\documentclass[
    aps,
    physrev,
    superscriptaddress,
    twocolumn,nofootinbib,
    amsmath,amssymb
]{revtex4-2}

\usepackage{physics}
\usepackage{graphicx}
\usepackage{bm}
\usepackage{physics}
\usepackage{color}
\usepackage[dvipsnames]{xcolor}
\usepackage{hyperref}
\usepackage{xurl}

\def\dd{\mbox{d}}
\def\br{\mathbf{r}}

\hypersetup{breaklinks=true}
\hypersetup{
    colorlinks,
    linkcolor={red!50!black},
    citecolor={blue!50!black},
    urlcolor={blue!80!black}
}

\begin{document}

\title{Simulating Quantum Turbulence with Matrix Product States}

\author{Felipe Gómez-Lozada}
\email[]{feg46@pitt.edu}
\affiliation{Department of Mechanical Engineering and Materials Science, University of Pittsburgh, Pittsburgh, Pennsylvania 15261, USA}

\author{Nicolas Perico-García}
\affiliation{Department of Physics and Astronomy, University of Pittsburgh, Pittsburgh, Pennsylvania 15261, USA}

\author{Nikita Gourianov}
\affiliation{Datalogisk institut, Københavns Universitet, Universitetsparken 1, 2100 København, Denmark}

\author{Hayder Salman}
\affiliation{School of Engineering, Mathematics, and Physics, University of East Anglia, Norwich Research Park, Norwich NR4 7TJ, United Kingdom}
\affiliation{Centre for Photonics and Quantum Science, University of East Anglia, Norwich Research Park, Norwich NR4 7TJ, United Kingdom}

\author{Juan José Mendoza-Arenas}
\affiliation{Department of Mechanical Engineering and Materials Science, University of Pittsburgh, Pittsburgh, Pennsylvania 15261, USA}
\affiliation{Department of Physics and Astronomy, University of Pittsburgh, Pittsburgh, Pennsylvania 15261, USA}

\date{\today}

\begin{abstract}

Quantum turbulence spans length scales from the system size $L$ to the healing length $\xi$, making direct numerical simulations (DNS) of the Gross-Pitaevskii (GP) equation computationally expensive when $L \gg \xi$. We present a matrix product state (MPS) solver for the GP equation that efficiently compresses the wavefunction by truncating weak interlength-scale correlations. This approach reduces memory usage by factors ranging from 10x to over 10,000x compared to DNS. We benchmark our approach on nonlinear excitations, namely dark solitons (1D) and quantized vortices (2D, 3D), capturing key dynamics such as Kelvin wave propagation and vortex ring emission in the case of vortex line reconnection. For turbulent states composed of multiple nonlinear excitations, we find that the memory compression of the MPS representation is directly proportional to the soliton or vortex densities. We also accurately reproduce established results from two-point correlation functions and energy spectra, where we recover the incompressible kinetic energy spectrum with little memory overhead. These results demonstrate the representative capabilities of the MPS ansatz for quantum turbulence and pave the way for studying this nonequilibrium state using previously-prohibited system sizes to uncover novel physics.

\end{abstract}

\maketitle

\section{Introduction}

The advent of the \emph{Bose-Einstein condensates} (BECs) in cold atom experiments in 1995~\cite{Anderson_Sci1995, Davis_PRL1995} is a milestone of many-body quantum phenomena. The striking properties of this state of matter, such as superfluidity, nonlinear dynamics, quantization of vortex circulation, and emergence of second sound~\cite{Pethick_2008, Paoletti_ARCMP2011, Madeira_ARCMP2020, Proukakis_CP2025}, have motivated a great deal of research. Moreover, the high tunability of cold-atom experiments~\cite{Dalfovo_RMP1999, Bloch_RMP2008, Bloch_NatP2012} means that these systems hold promise in a wide range of applications, including quantum metrology~\cite{Esteve_Nat2008, Ngo_SciR2021, Mao_NatP2023}, interferometry~\cite{Wales_CP2020, Grimshaw_PRL2022}, and quantum simulators~\cite{Bloch_NatP2012}. Nowadays, BECs have been realized in diverse unconventional quantum fluid systems~\cite{Bennemann_2013, Bennemann_2014, Proukakis_CP2025}. Seminal examples are exciton-polariton, magnon, and dipolar molecule condensates~\cite{Demokritov_Nat2006, Klaers_Nat2010, Altfeder_SciR2017, Bigagli_Nat2024}. As a result, such systems are ideal for studying fundamental nonequilibrium physics under controlled experimental conditions. This includes nonequilibrium thermal~\cite{Lamporesi_NatP2013} and quantum phase transitions~\cite{Yi_PRL2020}, dynamics of Bose-gas mixtures such as spinor condensates~\cite{Stamper-Kurn_RMP2013}, and quantum turbulence in BECs~\cite{White_PNAS2014, Navon_Nat2016, Fischer_PRA2025}.  Moreover, the ability to realize BECs in low-dimensional systems~\cite{Kinoshita_Nat2006, Langen_Sci2015} allows the study of 1D physics where integrable quantum field theories become relevant (e.g., in generalized hydrodynamics~\cite{Ruggiero_PRL2020, Bonnemain_JPA2022, Doyon_PRX2025}), as well as 2D quantum turbulence~\cite{Neely_PRL2013, Johnstone_Sci2019, Gauthier_Sci2019, Galka_PRL2022, Reeves_PRX2022, Karailiev_PRL2024} and Berezinskii–Kosterlitz–Thouless physics~\cite{Hadzibabic_Nat2006, Sunami_Sci2023}.

In modeling quantum fluids, much of the physics referred to in the above examples is dictated by a BEC well represented by a macroscopic occupation of a single-particle state~\cite{Pitaevskii_2016, Pethick_2008}. A natural starting point in modeling this physics is through a mean-field or classical-field approximation, given by the \emph{Gross-Pitaevskii} (GP) equation~\cite{Pethick_2008}. This model captures the nonequilibrium spatio-temporal dynamics of superfluids and BECs, including nonlinear excitations such as quantized vortices. In the context of quantum turbulence for BECs, the GP model accounts for compressibility in the system, enabling the coexistence and energy exchange between phonons and vortices. This mediates processes such as vortex dipole annihilation in 2D, vortex reconnections in 3D, dissipative vortex dynamics, and phonon scattering of vortices~\cite{Pismen_1999}. These mechanisms generate multiscale turbulent regimes~\cite{Barenghi_AVSQS2023}, from wave turbulence dominated by compressible modes~\cite{Kolmakov_PNAS2014} to Vinen or Kolmogorov turbulence where vortices dominate the dynamics~\cite{Barenghi_PNAS2014}.

The different studies mentioned above provide key insights into fundamental processes at prescribed scales. Therefore, extending GP-type simulations to span a broader spectrum of length scales is necessary for a deeper understanding of these diverse phenomena. Nevertheless, direct numerical simulations (DNS) for this endeavor are prohibitively costly when the system size $L$ is much larger than the healing length $\xi$ ($\sim$ size of a vortex) due to the number of computational variables growing very rapidly. A similar problem occurs for quantum many-body systems. For the latter, a challenge in their modeling corresponds to accurately representing the quantum state described by the many-body wavefunction that resides in an exponentially large-dimensional Hilbert space. However, since the many-body Schrödinger equation governs the wavefunction, it contains a significant amount of structure that a direct numerical solution does not exploit. \emph{Tensor networks} target the underlying structure by recognizing that, for many systems of interest, the governing Hamiltonian leads to limited entanglement within the many-body wavefunction~\cite{Schollwock_AP2011}. By decomposing the extensive quantum states into a series of low-rank tensors connected by virtual bonds, tensor networks can lead to a drastic reduction in the number of parameters that are required to describe the state of the system accurately~\cite{Orus_AP2014}.

The quantics representation of continuous variable functions leverages the analogy between many-body quantum systems and the solution of partial differential equations (PDEs) that resolve a large number of length scales. It does so by encoding real-space functions into quantum states of multiple qubits, where each two-level component represents a particular length scale in real space. Then, a tensor network decomposition is used to compress the given state, depending on the correlation distribution among length scales~\cite{Oseledets_SIAMMAA2010, Khoromskij_CA2011}. In particular, the hallmark matrix product state (MPS)~\cite{Orus_AP2014} (also called tensor train~\cite{Oseledets_SIAMJSC2011}), has been widely used for these purposes. Efficient MPS encoding of real-space functions has opened the door for using tensor network methods in the solution of large-scale PDEs. By exploiting the interlength-scale correlation structure, it has been possible to achieve low-rank representations of classical turbulent flows~\cite{Gourianov_NatComputSci2022, Gourianov_2022, Kornev_Mathematics2024, Kiffner_PRF2023, Peddinti_CP2024, Holscher_PRR2025, Gourianov_SciA2025, Adak_JSC2025, Pisoni_2025, Arenstein_2025}, plasma states~\cite{Ye_PRE2022, Ye_JPP2024}, wave phenomena~\cite{Fraschini_2024, Lively_2025} and different properties of continuous quantum systems~\cite{Shinaoka_PRX2023, Ritter_PRL2024, Rohshap_PRR2025}.

In this work, we develop a tensor network algorithm to simulate the nonlinear GP equation. We perform a systematic study of the quantics MPS for nonlinear excitations of the GP equation, namely dark solitons (1D) and vortices (2D and 3D). By using different examples of single and multiple nonlinear excitations, the latter corresponding to quantum turbulence cases, we explore the interlength-scale correlation structure of the states. That way, we characterize instances where the MPS representation reduces the memory usage with respect to the requirements of DNS. The main objective is to achieve low-rank representations of quantum turbulence simulations that can lead in the future to the discovery of new physics previously unattainable with classical methods. 

The manuscript is structured as follows. In Sec.~\ref{sec:Model}, we present the details of the GP equation along with its extension to include thermal effects. Then we introduce the tensor network and quantics frameworks in Sec.~\ref{sec:TN-QTT}, emphasizing the encoding of multidimensional real-space solutions of the GP equation and our time evolution method. In Sec.~\ref{sec:ElementaryExcitations}, we explore the use of MPS methods to tackle single nonlinear excitation problems produced by phase differences (dark solitons and vortices). Based on the latter, in Sec.~\ref{sec:QuantumTurbulence} we consider integrable and non-integrable quantum turbulence. This analysis includes: (1) a soliton gas in 1D that is also closely related to the problem of quantum generalized hydrodynamics~\cite{Suret_PRE2024};  (2) quantum turbulence in 2D comprising a random distribution of vortices and anti-vortices~\cite{Reeves_PRA2012, Nowak_PRA2012, Simula_PRL2014, Salman_PRA2016}; (3) a vortex tangle of a BEC in 3D that has been considered in other studies to uncover the similarities and differences between quantum and classical turbulence~\cite{Nore_PF1997, Berloff_PRA2002, Nowak_PRB2011}. Here, we characterize the compression capabilities of the MPS as a function of the system size and number of nonlinear excitations. Finally, in Sec.~\ref{sec:Conclusions} we present our closing remarks.

\section{Model} \label{sec:Model}

We simulate Bose gases governed by the damped GP equation~\cite{Choi_PRA1998, Tsubota_PRA2002, Bradley_PRX2012, Reeves_PRA2012, Reeves_PRL2013, Billam_PRL2014, Billam_PRA2015, Baggaley_PRA2018}. This equation is given by
\begin{align}
     i \hbar \frac{\partial \psi (\vb{r})}{\partial t} &= (1-i\gamma) \left[ -\frac{\hbar^2}{2m} \nabla^2 + V(\vb{r}) + g |\psi|^2 - \mu \right] \psi \, . \label{eqn:dGPE}
\end{align}
Here, $\psi$ is a macroscopic wavefunction, $m$ is the mass of the atomic species, $V(\vb{r})$ corresponds to a time-independent trapping potential, $g$ is the interaction coefficient that can be expressed in terms of the s-wave scattering length (and depends on the dimensionality of the system), and $\mu$ is 
the chemical potential. 

When $\gamma = 0$, we recover the conservative form of the GP equation corresponding to zero temperature. This form conserves the energy of the system as well as the total number of particles, $N_{\text{p}}=\int |\psi|^2 \dd^d \vb{r}$, where $d$ is the number of spatial dimensions. The GP equation provides a mean-field description of a Bose gas in which the macroscopic wavefunction is associated with the formation of a condensate. It can also be recovered from the many-body description of the Bose gas as written in second-quantized form. When the modes to be modelled are macroscopically occupied, it is common to replace creation and annihilation operators with commuting classical fields. Therefore, this is also referred to as the classical fields approximation and forms the basis for modeling Bose gases at finite temperature. 

A comprehensive modeling of finite temperature effects typically involves the stochastic GP equation coupled to a thermal bath~\cite{Rooney_PRA2012}. However, regimes exist where the dissipative effects of the noncondensed thermal component on the condensate can be well captured using the damped GP equation and where the damping coefficient, $\gamma$, is taken to be a function of temperature~\cite{Gauthier_Sci2019, Reeves_PRX2022}. Therefore, we use this model as the simplest applicable model capable of reproducing phenomena studied in cold atom BEC experiments. Unlike the conservative form of the GP equation, the damped model does not conserve energy. However, the average number of particles is conserved due to the inclusion of the chemical potential $\mu$ in Eq.~\eqref{eqn:dGPE}

Following standard practice in simulating properties of soliton gases and quantum turbulence, we set $V(\vb{r})=0$ and employ a periodic domain of extent $L^d$~\cite{Suret_PRE2024, Baggaley_PRA2018, Stagg_PRA2016}. In general, the interaction coefficient can be positive or negative since the scattering length can be experimentally tuned using Feshbach resonances~\cite{Chin_RMP2010}. In this work, we focus exclusively on the case with repulsive interactions, $g>0$, which ensures stability of the condensate even in high dimensions. Finally, we adopt a nondimensional form of Eq.~\eqref{eqn:dGPE} by working in units of the healing length, $\xi = \hbar / \sqrt{m\mu}$, and the speed of sound, $c = \sqrt{\mu/m}$. We also normalize the density of particles, $\lambda = N_{\text{p}} / L^{d}$, to unity. On the other hand, the damping term is typically taken to lie in the range $\gamma = 10^{-3}$ to $\gamma = 10^{-2}$~\cite{Choi_PRA1998, Tsubota_PRA2002, Bradley_PRX2012, Reeves_PRA2012, Reeves_PRL2013, Billam_PRL2014, Billam_PRA2015, Baggaley_PRA2018, Baggaley_PRE2011}. Therefore, it represents a small effect in comparison to other terms appearing in the equation.

\section{Tensor networks and quantics representation} \label{sec:TN-QTT}

\subsection{MPS encoding of continuous functions} \label{sec:EncodingMPS}

\begin{figure}[t]
    \centering
    \includegraphics[width=\linewidth]{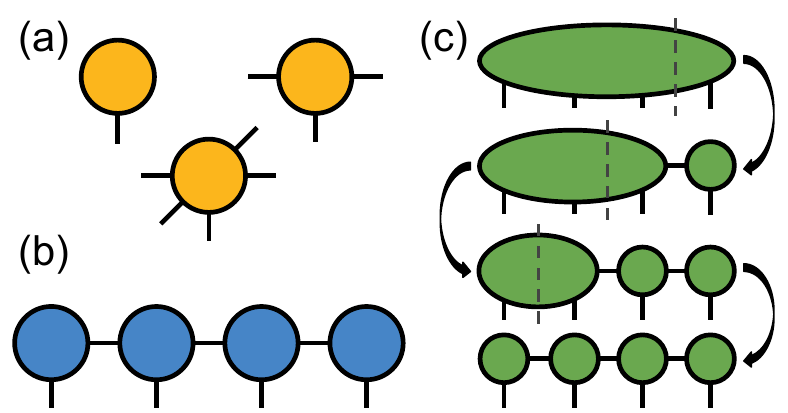}
    \caption{Examples of tensor network diagrams. (a) Tensors of rank one, three, and five. (b) Example of an MPS. (c) Creation of an MPS from a 4-rank tensor. Each dashed line corresponds to an SVD that splits the corresponding tensor.}
    \label{fig:TensorNetworkExample}
\end{figure}

Tensor networks consist of sets of tensors, which are typically illustrated graphically using geometric shapes. Lines attached to each shape represent the indices of each tensor; the number of lines represents the rank of the corresponding tensor, as shown in Fig.~\ref{fig:TensorNetworkExample}(a).
On top of these tensors, a topology is defined in which two joint indices (bonds) represent a tensor contraction without explicitly operating. A simple example of such a tensor architecture is shown in Fig.~\ref{fig:TensorNetworkExample}(b). This one-dimensional network is known as an MPS. In general, a tensor of arbitrary rank $N$ can be decomposed into an MPS by a sequence of singular value decompositions (SVDs), as depicted in Fig.~\ref{fig:TensorNetworkExample}(c) for $N = 4$. Assuming all free (noncontracted) indices are of dimension $D$, the iterative SVDs produces an MPS with bond dimensions $\chi_l = \min \left( D^l, D^{N-l} \right)$ for the $l$-th bond~\cite{Schollwock_AP2011}. The resulting singular values from the SVDs, named $S_{j,l}$ for the $j$-th singular value in decreasing order at the $l$-th bond after normalizing as $\sum_j S_{j,l}^2 = 1$, represent a measure of the correlations across each bond. The decay of these values gives an insight into which can be discarded while retaining the principal characteristic of the original state. The resulting truncation can either be done by setting a maximum bond dimension $\chi_{\text{max}}$ or a tolerance $\varepsilon$. In the first case, only the $j \leq \chi_{\text{max}}$ singular values are kept for all bonds. On the other, the bond dimension $\chi_l$ for the $l$-th bond is choosen such that $\sum_{j>\chi_l} S_{j,l}^2 \approx \varepsilon$. In this work, we use both forms of truncation and clarify which one is being used in each case. In particular, when we fix $\chi_{\text{max}}$, we also set $\varepsilon = 10^{-16}$ if the tolerance is not specified, which is a value close to machine precision set to eliminate singular values unresolved by the computer resolution.

Of particular importance are tensors whose MPS representations allow for low bond dimensions $\chi \ll D^{N/2}$ since the use of the MPS corresponds to a considerable reduction of memory. Here, the associated memory of an MPS $\ket{\psi}$ with $N$ free indices of dimension $D$ and a given set of bond dimensions $\chi_l$ is defined as the ratio of the number of parameters stored by the MPS to the parameters of the original tensor ($D^{N}$):
\begin{align} \label{eqn:memory}
    \text{Memory}(\ket{\psi}) = \frac{1}{D^{N}} \sum_{l=0}^{N} D\chi_l\chi_{l+1},
\end{align}
where we define $\chi_0 = \chi_{N} = 1$ for the missing edge bonds.
%\footnote{The memory measurement can be greater than one for large bond dimensions, given that the MPS representation is not unique as arbitrary unitary matrices can be added on any bond~\cite{Schollwock_AP2011}.}
This insight has been widely used for quantum many-body gapped systems in 1D and even in higher dimensions, allowing the study of previously unattainable quantum states because of the exponentially large Hilbert space~\cite{Verstraete_AdvP2008, Orus_AP2014}.

To leverage tensor networks for simulating PDEs, we first map real-space functions into the quantum state of a many-qubit system for its subsequent decomposition using MPS. We illustrate the encoding for functions defined over a one-dimensional domain ($d=1$); this process can be directly extended to higher dimensions. Given a function $f(x)$ with $x \in [0,L)$ we define a grid of $2^N$ points for a positive integer $N$ as $x_m = L \cdot m/2^N$ with $m = 0, 1, \dots 2^N-1$. We adopt the quantics representation~\cite{Oseledets_SIAMMAA2010, Khoromskij_CA2011} where the variable $x_m$ is written in its binary form 
\begin{align} \label{eqn:QTT}
    x(n_1,n_2,\dots,n_N) = L\sum_{l=1}^N \frac{n_l}{2^l}, \quad n_l = 0, 1.
\end{align}
The latter defines a rank-$N$ tensor with indices of dimension $D = 2$ representing the continuous function in the grid as $f_{n_1,n_2,\dots,n_N} = f(x(n_1,n_2,\dots,n_N))$. This definition corresponds to encoding the function in the quantum state of an $N$-qubit system, where the $l$-th qubit represents the $L \cdot 2^{-l}$ length scale of the system. With this encoding, we obtain a low-rank MPS using iterative SVDs, as shown in Fig.~\ref{fig:TensorNetworkExample}(c), and discard singular values with the least contribution to retain the principal details of the original tensor. The previous process is possible if the correlations across the MPS are bounded. The latter is known to occur, for example, in 1D quantum many-body systems that satisfy the area law. These quantum systems exhibit an exponential memory reduction of that which is required by the original state~\cite {Schollwock_AP2011}. For continuous functions in the quantics MPS ansatz, this corresponds to bounded correlations among length scales. This correlation structure is a phenomenon characteristic of multiscale physics and specifically of interest in turbulent flows due to the Richardson cascade. The latter was first exploited in Ref.~\cite{Gourianov_NatComputSci2022, Gourianov_2022} to obtain large memory compressions with the quantics MPS framework for modeling classical turbulence.

\begin{figure}[t]
    \centering
    \includegraphics[width=\linewidth]{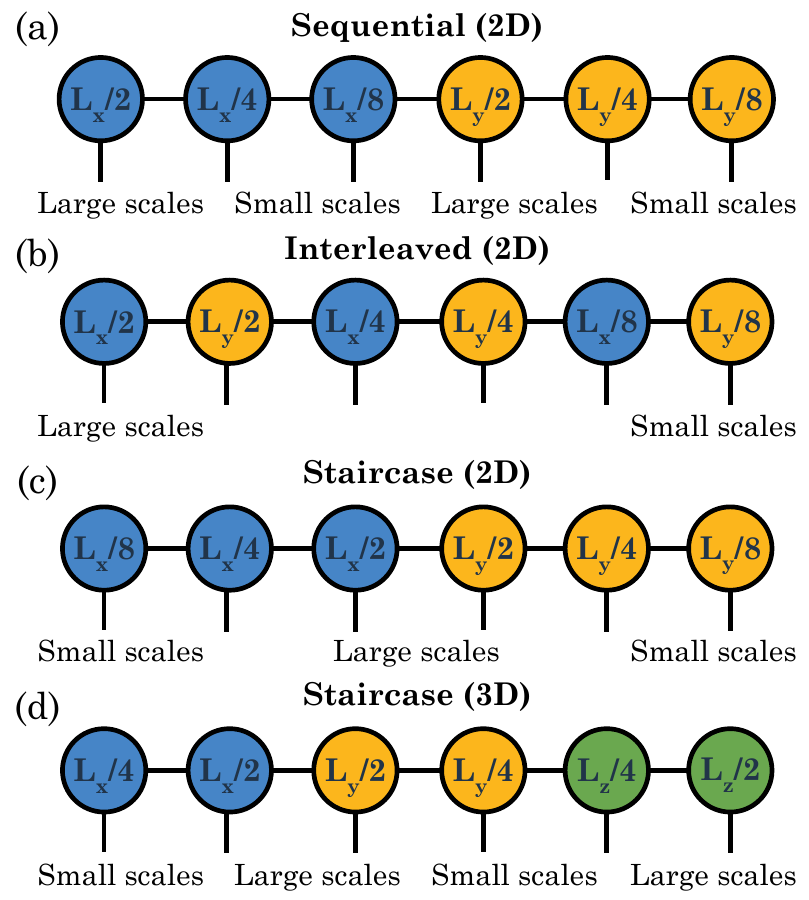}
    \caption{Example of MPS orderings of 2D (3D) functions in a box of size $L_x \times L_y$ ($\times L_z$) with 3 (2) qubits per axis. Each tensor represents a length scale of the corresponding variable (blue for $x$, orange for $y$, and green for $z$). (a) Sequential order in 2D stacks the qubits of the same variable in decreasing length scale order. (b) Interleaved order in 2D couples qubits at the same length scale. (c) Staircase order in 2D, analogous to (a), with the largest length scales coupled at the middle of the chain. (d) Staircase order in 3D extends the 2D case by coupling the last axis through the smallest length scale.}
    \label{fig:MPSOrder}
\end{figure}

For higher dimensions $d > 1$, an analogue iterative process can be performed using $dN$ qubits\footnote{Each spatial dimension can have a different number of qubits, resulting in different grid resolutions. Here we always assume the same resolution for all spatial dimensions in the lattice.}. There are multiple ways to order the qubits of each axis and length scales into a one-dimensional MPS structure. The sequential order shown in Fig.~\ref{fig:MPSOrder}(a) is a direct analogy to the 1D case. 
The interleaved or alternating order corresponds to locating the qubits of the same scale next to each other, as depicted in Fig.~\ref{fig:MPSOrder}(b). Here we argue that extending the sequential ordering such that the interaxis coupling is done through the same scale is beneficial for an efficient (e.g., accurate and highly compressed) MPS representation; we refer to this encoding as the staircase ordering, illustrated in Fig.~\ref{fig:MPSOrder}(c). For the staircase encoding in 3D, indicated in Fig.~\ref{fig:MPSOrder}(d), we couple the last spatial dimension through the remaining length scale available. This ordering is also known as the ``sequential, interleave'' order~\cite{Ye_JPP2024}. The 3D staircase requires the specification of the ordering of the axes ($x$, $y$, or $z$) across the MPS, resulting in six different possible permutations. We compare different MPS ordering to obtain the low-rank decomposition with the best compression for our simulations.

\subsection{Time evolution method} \label{sec:TimeEvolution}

We solve the damped GP equation with the DNS and the MPS method, to subsequently compare them to characterize the accuracy of the latter. For the DNS, it is common to use the split-step Fourier method~\cite{Bao_JCP2003}, which relies on the decomposition of Eq.~\eqref{eqn:dGPE} into two sets of operators, namely $\hat{U} = |\psi|^2$ and $\hat{K} = -\frac{1}{2}\nabla^2$ (after introducing natural units). By using Strang splitting and the fast Fourier transform (FFT), the method applies the evolution operators of $\hat{U}$ and $\hat{K}$ in position and momentum space, respectively, where they are diagonal. We implement the second-order method with the evolution operator
\begin{align} \label{eqn:strang}
    e^{-i \Delta\tau (\hat{K} + \hat{U})} = e^{-i(\Delta\tau/2) \hat{K} } e^{-i\Delta\tau \hat{U} } e^{-i(\Delta\tau/2) \hat{K}} + \mathcal{O}(\Delta\tau^2),
\end{align}
where $\Delta\tau = (1-i\gamma)\Delta t$, $\Delta t$ represents a discretisation timestep, and FFTs are applied before and after evolving the system with the term corresponding to the kinetic energy. This results in a computational cost scaling as $\mathcal{O}(M\log M)$ where $M=2^{Nd}$ is the total number of grid points. If $\hat{U}$ and $\hat{K}$ are interchanged in Eq.~\eqref{eqn:strang}, the evolution operator is not symmetric with respect to the nonlinear term, which degrades the accuracy.

To evolve an MPS according to Eq.~\eqref{eqn:dGPE}, we rely on the time-dependent variational principle (TDVP)~\cite{Haegeman_PRL2011, Haegeman_PRB2016}. In analogy to the celebrated density matrix renormalization group (DMRG)~\cite{White_PRL1992}, a sweeping strategy is implemented where each tensor is evolved locally using a projection operator~\cite{Haegeman_PRB2016}. The projector ensures the state remains in the MPS manifold of a given bond dimension. Completing a sweep through the system corresponds to evolving the system one timestep~\cite{Paeckel_AP2019}. In particular, we use the two-site TDVP formulation~\cite{Paeckel_AP2019}, which allows us to vary the bond dimensions of the MPS during the time integration of our equations. The TDVP has shown promising results for the evolution of long-range interacting systems~\cite{Yang_PRB2020}. This algorithm has also been used for solving PDEs~\cite{Ye_JPP2024, Connor_2025}, where it can be numerically more stable than using DNS~\cite{Ye_JPP2024}.

To apply the TDVP to our nonlinear damped GP equation, we adopt the same Strang splitting presented in Eq.~\eqref{eqn:strang} and evolve the state using the TDVP with the corresponding operators $\hat{K}$ and $\hat{U}$. The construction of these operators in the quantics MPS representation is detailed in Appendix~\ref{sec:OperationsQTT}, where we use eighth-order central finite differences represented in real space for the derivatives. The computational costs for TDVP are $\mathcal{O}(dN\eta \chi^3)$ with $\eta$ the bond dimension of the operator that drives the evolution and $\chi$ the bond dimension of the MPS being evolved. For the kinetic term, we have $\eta \sim \mathcal{O}(1)$ while for the interaction term $\eta \sim \chi$ due to the nonlinearity (see Appendix~\ref{sec:OperationsQTT}). Using these operators for TDVP corresponds to scalings of $\mathcal{O}(dN\chi^3)$ and $\mathcal{O}(dN\chi^4)$, respectively, the latter being the most costly computational step of the method.

We develop our code using the Julia programming language~\cite{Bezanson_SIAMR2017}. For the split-step Fourier method (DNS), we use the FFTW package~\cite{Frigo_PIEEE2005}. Our MPS simulations are based on the TDVP implementations of the ITensor library~\cite{Fishman_SciPPC2022, Fishman_SciPPC2022a}. 
%The code for the simulations can be found in \fg{ref}.
The simulations are performed in serial with an AMD EPYC 9575F processor. We use GPU acceleration supported by the ITensor library~\cite{Fishman_SciPPC2022, Fishman_SciPPC2022a} for the most computationally demanding calculations\footnote{This is only useful for large bond dimensions; if this value is low, the GPU simulation has an overhead compared to the serial implementation.}. In those cases, we use the NVIDIA A100 with 40GB of memory in conjunction with an AMD EPYC 7742 processor.

\section{Nonlinear excitations}\label{sec:ElementaryExcitations}

To demonstrate that MPS successfully captures the physics governed by the GP equation, we first assess the performance of the TDVP-based evolution of nonlinear excitations produced by phase jumps. Namely, we simulate dark solitons in 1D, vortex dipoles in 2D, and vortex rings in 3D, and compare with the corresponding DNS. Furthermore, we evidence the significant advantage of MPS over DNS by simulating vortex line reconnections with substantial memory compression.

\subsection{1D Dark Soliton}\label{sec:DarkSoliton}

As a first test case, we study the evolution of a dark soliton, i.e., a density hole in the condensate which moves at a constant speed. The wavefunction is defined in a box of size $L = 32 \, \xi$ such that the soliton is initially located at the center of the condensate with a translational speed of $v \approx 0.19\,c$. For this, we use the initial condition from Ref.~\cite{Sato_NJP2016}, detailed in Appendix~\ref{sec:DarkSoliton_Initial}. The soliton is encoded in a quantics MPS with $7$ qubits, corresponding to a grid of $2^7 = 128$ points which resolves up to a quarter of the healing length ($\xi/4$). We have verified that this resolution is enough to reach convergence in the DNS. We use iterative SVDs to create the corresponding MPS for a variable $\chi_{\text{max}}$ as explained in Sec.~\ref{sec:EncodingMPS}.

We evolve the system using both the DNS and MPS methods until the soliton performs a complete lap across the box. Afterwards, we compare the solution with the analytical result from Appendix~\ref{sec:DarkSoliton_Initial} using the infidelity measure
\begin{align} \label{eqn:Infd}
    I(\ket{\psi}, \ket{\phi}) = 1 -  \frac{\left|\braket{\psi}{\phi}\right|}{\sqrt{\braket{\psi}\braket{\phi}}},
\end{align}
commonly used in quantum information theory~\cite{Nielsen_2010}. To perform the comparison, the MPS is contracted to obtain the underlying vector; then Eq.~\eqref{eqn:Infd} is used.

\begin{figure}[t]
    \centering
    \includegraphics[width=\linewidth]{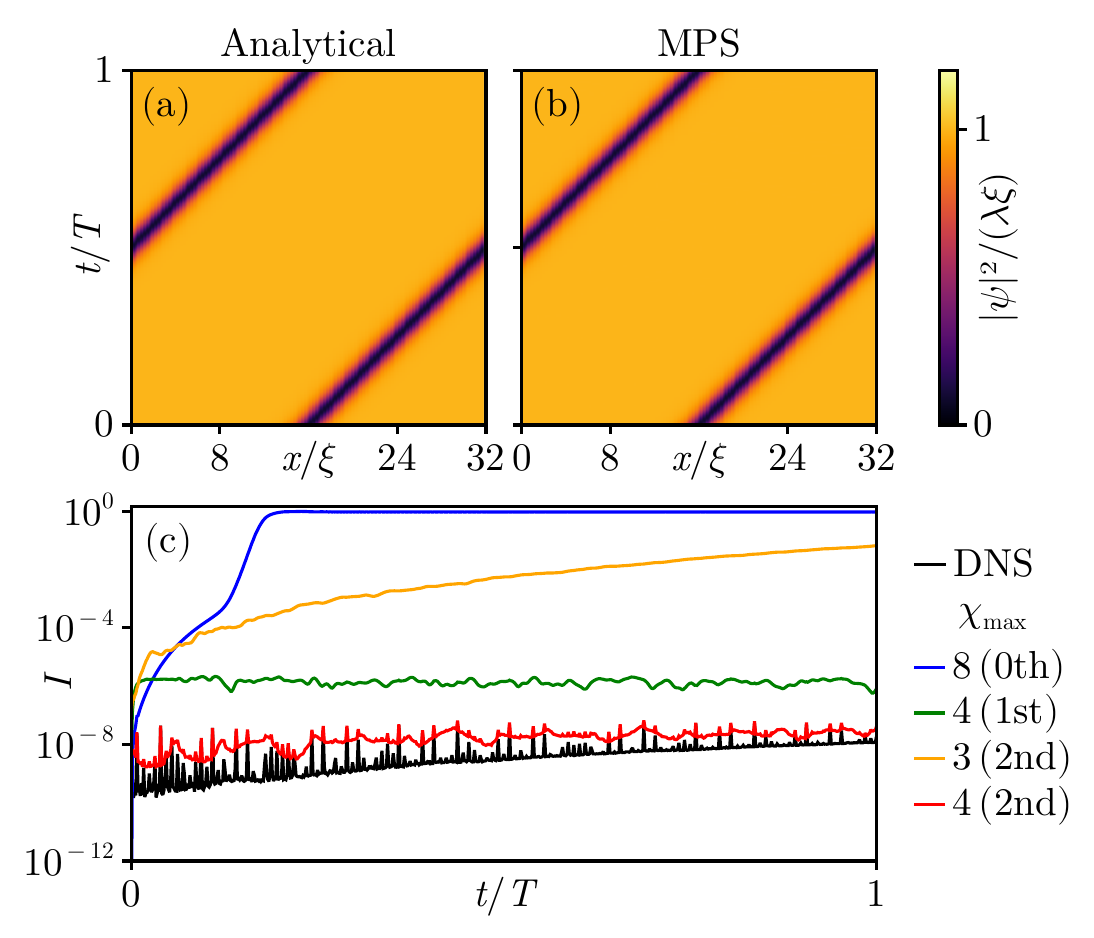}
    \caption{
        Evolution of a dark soliton. (a) Analytical density profile $|\psi|^2$ as a function of time $t$. (b) MPS-evolved density profile with $\chi_{\text{max}} = 4$. (c) Infidelities $I$ with respect to the analytical ansatz for the DNS and MPS method with different $\chi_{\text{max}}$ and splitting orders (indicated inside the parentheses). The soliton is located in a box of size $L = 32\,\xi$ with a velocity $v \approx 0.19\,c$. The time is scaled by the period $T = L/v \approx 165\,\xi/c$. We use $7$ qubits ($128$ points), the timestep is $\Delta t = 2^{-5}\,\xi/c$ and the evolution is undamped ($\gamma = 0$).
    }
    \label{fig:DarkSoliton}
\end{figure}

In Fig.~\ref{fig:DarkSoliton}(a), we show the density profile $|\psi|^2$ as a function of time $t$ for the exact evolution of the dark soliton. 
Next to that, in Fig.~\ref{fig:DarkSoliton}(b), we present the result of the simulation using the MPS method with $\chi_{\text{max}} = 4$. To the naked eye, both results look identical. The MPS dynamics shows the expected physical behavior: The soliton moves in a straight line in $x-t$ space (i.e., it has a constant speed) while preserving its shape.
This comparison indicates that the dynamics is well represented for $\chi_{\text{max}} = 4$, as highlighted by the infidelity plotted in Fig.~\ref{fig:DarkSoliton}(c) (red line) which saturates around $I \approx 10^{-8}$ and approaches the corresponding DNS infidelity (black line)\footnote{The oscillations are accumulated errors in the numerical procedure used to obtain the initial condition shown in Appendix~\ref{sec:DarkSoliton_Initial}.}. If we decrease the bond dimension to $\chi_{\text{max}} = 3$, the truncation error quickly increases, and the MPS simulation strongly deviates from the analytical solution. This results in a rapid growth of the infidelity towards the maximal value of $I \approx 1$ (orange line).

We also analyze the effect of splitting orders different from Eq.~\eqref{eqn:strang} for the evolution operator. If we only use a first-order splitting
\begin{align} \label{eqn:strang1}
    e^{-i\Delta\tau (\hat{K} + \hat{U})} = e^{-i\Delta\tau \hat{K} } e^{-i\Delta\tau \hat{U} } + \mathcal{O}(\Delta \tau),
\end{align}
with the same timestep and bond dimension $\chi_{\text{max}} = 4$ (green line in Fig.~\ref{fig:DarkSoliton}(c)), the infidelity is almost two orders of magnitude higher than its second-order counterpart. The second-order method only requires an additional TDVP evolution (using $\hat{K}$) to the first-order method. This additional calculation has a lower scaling ($\mathcal{O}(N\chi^3)$) than the nonlinear evolution ($\mathcal{O}(N\chi^4)$); hence, the upgrade to second-order splitting from first-order is computationally inexpensive compared to the rest of the algorithm. In contrast, for the DNS case, going from first to second splitting order doubles the computational resources as it requires an additional FFT. Higher orders would require more applications of the $\hat{U}$ operator, which would significantly increase the computational cost. This larger number of operations per timestep would also enhance the accumulation of truncation errors in the MPS evolution~\cite{Connor_2025}.

We can also avoid the splitting, i.e., calculate $\hat{K}+\hat{U}$ and evolve the state using TDVP accordingly. In that case, the simulation breaks down even for $\chi_{\text{max}} = 8$ as depicted by its infidelity $I \approx 1$ (blue line in Fig.~\ref{fig:DarkSoliton}(c)). We find that to achieve infidelities close to the first order result ($I \approx 10^{-6}$) without splitting, this simulation requires a reduction of the timestep by a factor of $\approx 512$. Hence, splitting the operators is crucial for TDVP to capture efficiently the evolution of the GP equation.

\subsection{2D Vortex dipole} \label{sec:VortexDipole}

For 2D, we analyze the vortex dipole: a pairing of two vortices of opposite charge that move in a straight line at a constant speed due to their interaction. Using a Padé approximation ansatz, we create an initial vortex dipole with a fixed distance $d = 10\, \xi$ between the vortices (dipole length) in a system of size $L = 32 \, \xi$ with a resolution of $\xi/4$. An initial imaginary time evolution removes sound waves from the ansatz while keeping the dipole length approximately unaffected. We describe this process in Appendix~\ref{sec:VortexDipole_Initial}. 

Since we have more than one dimension, there are multiple ways to order the length scales across the MPS encoding, as shown in Fig.~\ref{fig:MPSOrder}. In our case, we choose the staircase ordering that minimizes the correlations among the large length scales (Fig.~\ref{fig:MPSOrder}(c)). This choice is reasonable because the vortex core structure is equal among all vortices; therefore, the structure of the wave function is mainly characterized by the vortex distribution, encoded in the large length scales of the MPS. A detailed analysis of the correlations is shown in Appendix~\ref{sec:Orderings2D}.

\begin{figure}[t]
    \centering
    \includegraphics[width=\linewidth]{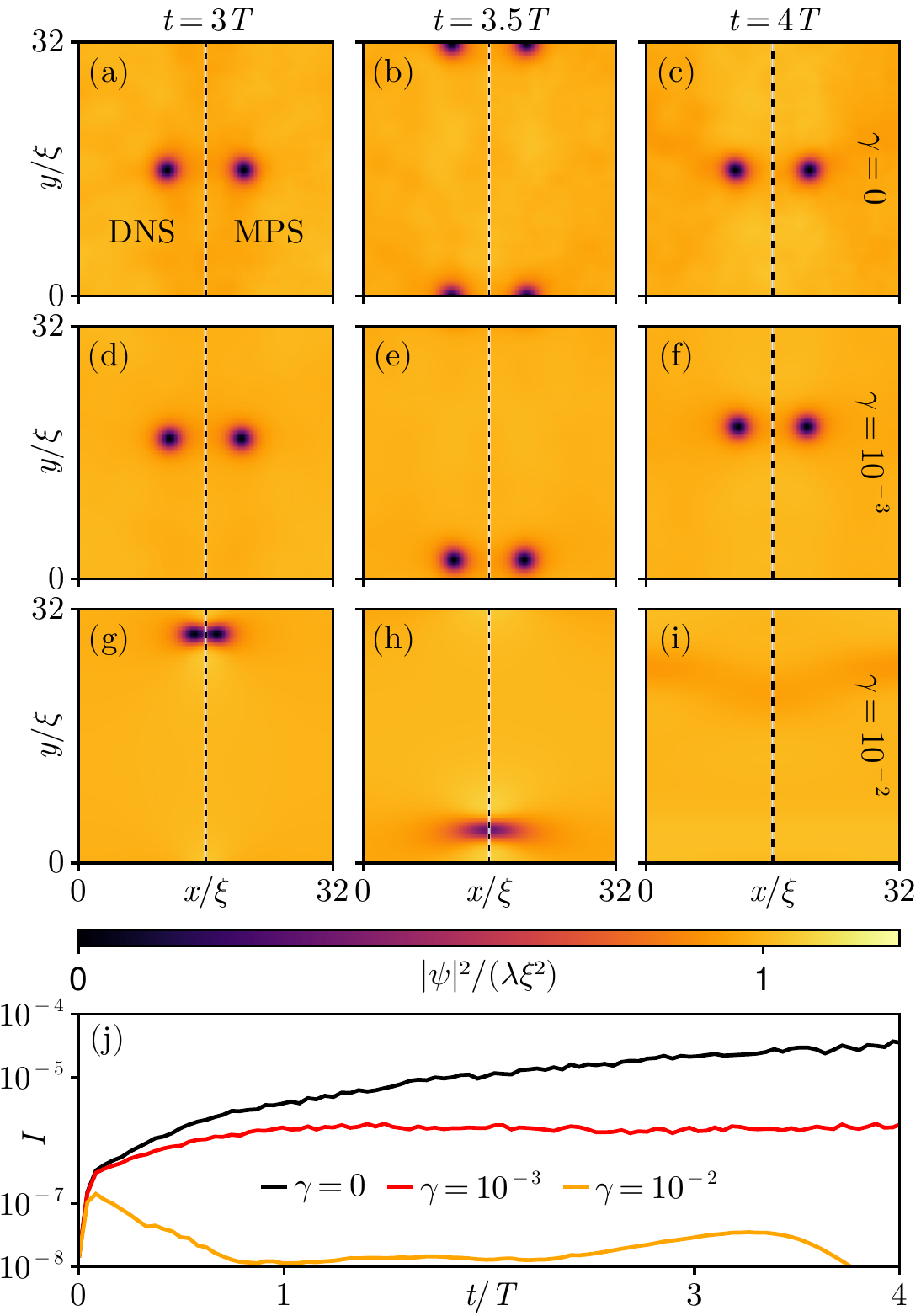}
    \caption{
        Evolution of a vortex dipole. (a-i) Density profile $|\psi|^2$ comparison at various times $t$ between DNS (left) and MPS with $\chi_{\text{max}} = 12$ (right) using damping coefficients of $\gamma = 0$ (a-c), $\gamma = 10^{-3}$ (d-f) and $\gamma = 10^{-2}$ (g-i). (j) Infidelity $I$ between DNS and MPS with previous truncation parameters and damping coefficients. The system size is $L = 32\,\xi$ with an initial vortex dipole $d \approx 10\,\xi$ after an imaginary time evolution evolution of $it = 1\,\xi/c$. Time is scaled by the period of the dipole to traverse the system size without damping ($\gamma = 0$), given by $T = 242\,\xi/c$. We use a grid of $7$ qubits per axis ($128 \times 128$ points) with $\Delta t = 2^{-6}\,\xi/c$. 
    }
    \label{fig:VortexDipoleEvolution}
\end{figure}

We let the vortex dipole evolve for four periods, with a period being the time that it takes the dipole to traverse the system size in the vertical direction\footnote{The period is measured without dissipation since it is the only case in which the dipole moves at a constant speed.}. In Figs.~\ref{fig:VortexDipoleEvolution}(a-c), we show density profiles at various times for $\gamma = 0$, comparing the DNS (left) and MPS (right) results with $\chi_{\text{max}} = 12$. 
We observe that both vortex dipole trajectories match, with the MPS using just $14\%$ of the memory required by DNS. For lower bond dimensions, the truncation error grows rapidly, resulting in perturbations to the vortex dipole that lead to unphysical results at long evolution times. In contrast to the soliton case (Fig.~\ref{fig:DarkSoliton}(c)), the corresponding infidelity with the DNS shown in Fig.~\ref{fig:VortexDipoleEvolution}(j) grows steadily during the simulation (black line), resulting in a final value of $I\approx 10^{-4}$. This increase is an effect of remnant sound waves released from the dipole at the initial timesteps, which accumulate in the system over time. An MPS does not well represent these sound waves since they are uncorrelated, hence the increase in the infidelity with respect to the DNS.

When we consider a system coupled to a thermal bath, modeled by the damped GP equation (coefficient $\gamma > 0$), the damping removes the background noise but also decreases the dipole length. In Figs.~\ref{fig:VortexDipoleEvolution}(d-f) we observe the evolution of the dipole for $\gamma = 10^{-3}$ where sound waves have already been anhihiated, and the damping is low enough that the dipole length is still $d \approx 10\,\xi$. The infidelity (red line in Fig.~\ref{fig:VortexDipoleEvolution}(j)) features a plateau similar to that observed in the soliton case (Fig.~\ref{fig:DarkSoliton}(c)) around $I \approx 10^{-6}$. For a larger damping rate of $\gamma = 10^{-2}$, the density profiles in Fig.~\ref{fig:VortexDipoleEvolution}(g-i) depict the vortex pair annihilating after three periods, decaying into sound waves. The MPS ansatz also captures this process at this truncation level ($\chi_{\text{max}} = 12$). The anihilation is indicated by a local maximum in the corresponding infidelity (orange line in Fig.~\ref{fig:VortexDipoleEvolution}(j)). After the decay of the vortex dipole, the infidelity decreases steadily, signaling the transition to the uncorrelated ground state. While compressible waves are harder to capture than the vortex structures, we note that this can be tackled by increasing the bond dimension. For instance, the undamped case $\gamma = 0$ can reach infidelities of $I \approx 10^{-6}$ after four periods by increasing $\chi_{\text{max}} = 16$. The larger bond dimension almost doubles the required computational memory, namely up to $23\%$ of that of DNS. Nevertheless, the simulation of the dipole does not require such levels of accuracy to capture the vortex dynamics.

\subsection{3D Vortex ring} \label{sec:VortexRing}

Next, we study the modeling of a BEC in 3D. Similarly to the previous sections, we begin by considering one of the most elementary vortex excitations that can arise in 3D, namely a vortex ring. This excitation moves at a constant speed along its axis of symmetry due to its self-induced velocity, in analogy to the vortex dipole from Section~\ref{sec:VortexDipole}. For the initial condition, we use the dipole ansatz and rotate it around the midpoint of the vortex pair to generate the desired ring such that its axi-symmetry is aligned with the $z$-axis. Then, we evolve the ring with imaginary time evolution to dampen sound waves. This construction is explained in Appendix~\ref{sec:VortexRing_Initial}. We use the staircase ordering for the length scales in the quantics MPS in extension to the 2D results from Sec.~\ref{sec:VortexDipole}. Within this MPS structure, we order the axes as $xyz$ (Fig.~\ref{fig:MPSOrder}(d)). Changing this choice does not appreciably affect the results as discussed in Appendix~\ref{sec:Orderings3D}.

\begin{figure}[t]
    \centering
    \includegraphics[width=\linewidth,trim={0.0cm 0.0cm 0.0cm 0.0cm},clip]{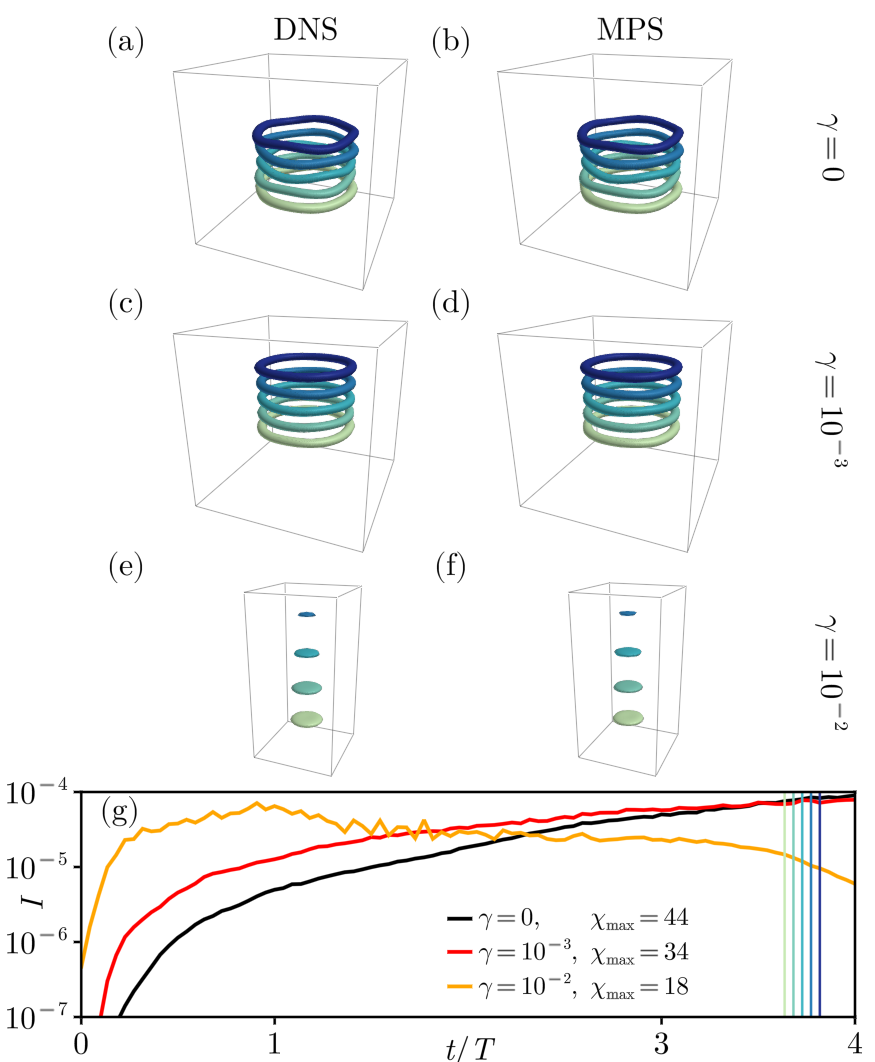}
    \caption{
        Evolution of a vortex ring. (a-f) Density contours at $|\psi|^2 = 0.1 \, \lambda \xi^3$, color-coded for different snapshot times $t$, using DNS (left) and MPS (right) methods and different damping coefficients: $\gamma = 0$ (a,b), $\gamma = 10^{-3}$ (c,d) and $\gamma = 10^{-2}$ (e,f). (g) Infidelity curves $I$ for each $\gamma$, indicating the corresponding $\chi_{\text{max}}$. The initial ring radius is $R \approx 5 \, \xi$ after an imaginary time evolution of $it = 1 \, (\xi/c)$, in a system size $L = 32\,\xi$ with a grid of $128^3$ points (7 qubits per axis). For DNS, we use $\Delta t = 2^{-7}\, \xi/c$ while for MPS $\Delta t = 2^{-6}\, \xi/c$. The density contours are shown inside boxes $[L/4, 3L/4]^2\times[0, L/2]$ (a-d) and $[3L/8, 5L/8]^2\times[L/2, L]$ (e,f). Time is scaled by the period $T = 88\, \xi/c$ that takes the undamped vortex ring to travel the box length $L$. Vertical lines in (g) indicate the snapshot times selected for (a-f). 
    }
    \label{fig:VortexRingEvolution}
\end{figure}

We analyze the dynamics of a vortex ring of radius $R = 5 \, \xi$ inside a box of size $L = 32 \, \xi$ in Fig.~\ref{fig:VortexRingEvolution}.
For that, we let the ring evolve until it travels the length of the domain $L$ four times (measured in the undamped limit $\gamma = 0$). Also, we set a grid resolution of $\xi/4$ in each spatial dimension. Then, we analyze the effect of different damping coefficients, $\gamma$, on the dynamics. In all cases, we choose the lowest $\chi_{\text{max}}$ for which the infidelity of the DNS and MPS is $I \lesssim 10^{-4}$. The zero-temperature case (panels (a,b)) requires the largest $\chi_{\text{max}} = 44$ of all the $\gamma$ values analyzed in the following, given the accumulation of sound waves in the system. Even in this case, we can accurately resolve the dynamics of the vortex ring using only $2\%$ of the memory that the DNS requires. This significant compression is further emphasized by the corresponding infidelity (black curve in panel (g)). Here, we note that we can simulate the MPS with a timestep twice as large as that of DNS. We find that increasing the timestep of the DNS leads to numerical instabilities for $\gamma = 0$, which severely affect the accuracy of the results. We argue that this stability property of the TDVP emerges because small numerical oscillations can be truncated during the evolution at each timestep, thus preventing their accumulation. A similar behavior was found in TDVP simulations for the Vlasov-Poisson equation~\cite{Ye_JPP2024}.

For non-zero $\gamma$ we require lower $\chi_{\text{max}}$ to achieve the same infidelity of $I \lesssim 10^{-4}$ (red and orange lines in Figs.~\ref{fig:VortexRingEvolution}(g)). Setting $\gamma = 10^{-3}$ (Figs.~\ref{fig:VortexRingEvolution}(c,d)) suppresses the sound waves, as signaled by the unperturbed circular structure of the vortex ring in contrast with the undamped simulation (Figs.~\ref{fig:VortexRingEvolution}(a,b)). In this case, due to the vanishing sound waves, we require $\chi_{\text{max}} = 34$ and $1\%$ of the memory that DNS uses. This is similar to the behavior of the vortex dipole as presented in Fig.~\ref{fig:VortexDipoleEvolution}(j). Increasing the damping to $\gamma = 10^{-2}$ (Figs.~\ref{fig:VortexRingEvolution}(e,f)) causes the vortex ring to shrink to a size of the order of the healing length and subsequently transitions into a rarefaction wave (the contour subsequently vanishes for the final timestep). The MPS well captures this process with $\chi_{\text{max}} = 18$. As a result, it uses $0.4\%$ of the memory required by DNS, corresponding to a reduction by a factor of $\mathcal{O}(10^3)$. This simple example clearly demonstrates the significant compression that can be realized with the MPS for 3D simulations.

\subsection{Vortex line reconnection}\label{sec:VortexLineReconnection}

We now demonstrate that the time-dependent MPS can accurately capture more complex dynamics. An important phenomenon associated with the physics of 3D quantum fluids is that of vortex line reconnections. Such events can induce a change of topology of the vortex lines whereby two line segments interchange ends upon reconnecting~\cite{Koplik_PRL1993, Serafini_PRX2017, Villois_PRL2020}. Here, we demonstrate how the quantics MPS framework can simulate this behavior with vast memory compression ratios.

\begin{figure}[t]
    \centering
    \includegraphics[width=\linewidth,trim={0.0cm 0.0cm 0.0cm 3.2cm},clip]{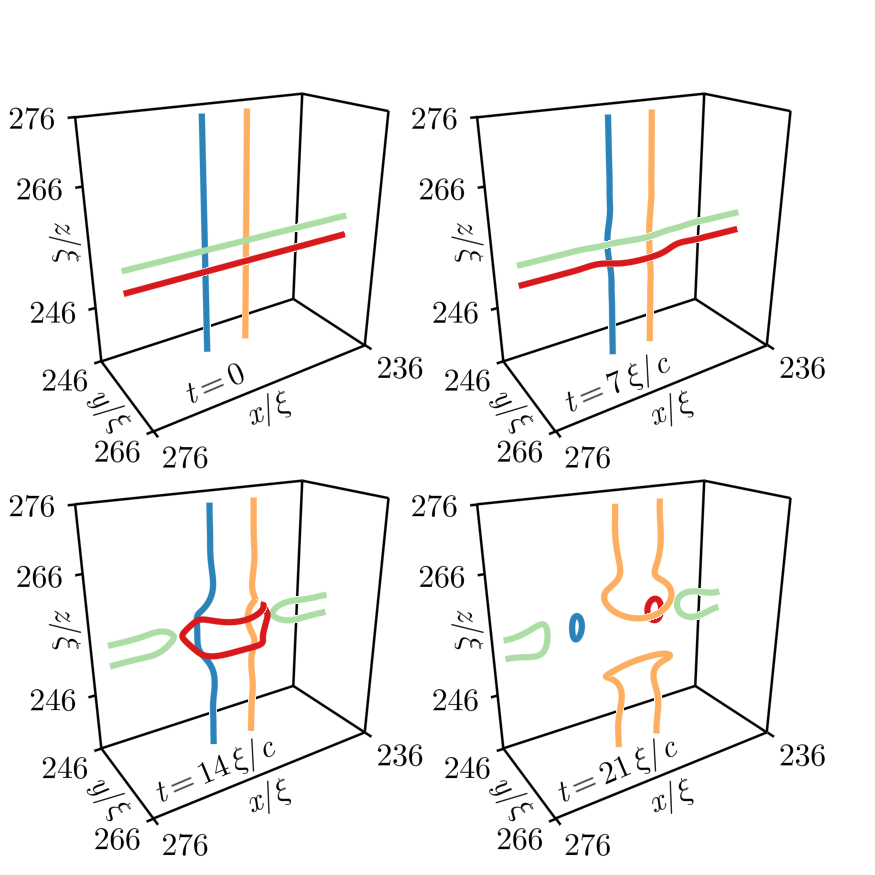}
    \caption{
        Vortex reconnection of perpendicular dipoles. We display vortex lines at different times $t$ and highlight the topology changes using different colors for disconnected lines. We use a tracking resolution $\zeta = \xi/4$ (see Appendix~\ref{sec:VortexTracking}) inside the box $[236\,\xi, 276\,\xi]\times[246\,\xi, 266\,\xi]\times[236\,\xi, 276\,\xi]$ with $L=512\,\xi$, $\chi_{\text{max}} = 80$ and $\Delta t = 2^{-6} \, \xi/c$. The dipoles have an initial intervortex separation of $d = 8\,\xi$ (vertical) and $d = 4\,\xi$ (horizontal), separated by a distance $l = 6\,\xi$. We apply an initial imaginary time evolution of $it = 1 \, (\xi/c)$.
    }
    \label{fig:VortexReconnectionInit}
\end{figure}

We study four vortex lines evolved via the GP equation with $\gamma = 0$: two pairs of counter-rotating vortices aligned along the $z$-axis ($x$-axis) with dipole lengths $d=8\,\xi$ ($d=4\,\xi$), separated by $l=6\,\xi$. The simulation utilizes a cubic domain of $L=512\,\xi$ to capture the reconnection dynamics and prevent unphysical sound-wave interference resulting from the periodic boundaries. Vortices are resolved with $\xi/4$ spacing (11 qubits per axis), resulting in a $2048^3 \approx 10^{10}$-point grid. Previous studies used grids of $\sim10^{6}$–$10^{7}$ points in smaller domains, applying different filters or boundary conditions to limit the unphysical sound-wave interference from reconnections~\cite{Kerr_PRL2011, Villois_PRF2017}. In our case, the quantics MPS allows us to work with a significantly larger domain size. To avoid the use of iterative SVDs for the creation of the initial condition, which is impractical for the grid sizes considered here, we extend the creation of quantics MPS for 2D dipoles to 3D via delta tensors. We detail this process in Appendix~\ref{sec:VortexLineReconnection_Initial}. Due to the large grid size we employed to model this physics, it is impractical to use DNS without resorting to highly involved parallel computing paradigms~\cite{Kobayashi_CPC2021}. For instance, a snapshot of the DNS requires a memory of $\sim 128 \, \text{GB}$. Then, instead of comparing to the respective DNS, alternative convergence criteria are discussed in Appendix~\ref{sec:ConvergenceReconnection}. These tests confirm that the results vary by less than $10^{-4}$ in infidelity when the bond dimension is increased. Also, the energy and linear momentum are conserved, as expected for the GP equation, up to relative errors of $10^{-6}$ and $10^{-2}$, respectively.

In Fig.~\ref{fig:VortexReconnectionInit}, we show the initial steps of the simulation in a close-up to the point of contact of the vortex lines. We note that to fully exploit the compression capabilities of the MPS, post-processing of the data necessitates techniques specifically tailored to the MPS representation. Otherwise, any advantage of the tensor network encoding would be lost if the MPS were contracted into the full vectorized representation. We overcome this challenge by adapting a vortex-tracking algorithm proposed in Ref.~\cite{Villois_JPA2016} to extract the vortex line data using a combination of MPS sampling and Fourier interpolation, which is detailed in Appendix~\ref{sec:VortexTracking}. With this tracking tool, we observe in Fig.~\ref{fig:VortexReconnectionInit} how the vortex lines reconnect into four U-shaped lines, one for each direction of the original dipoles, along with remnant vortex rings which are emitted following the reconnection~\cite{Kerr_PRL2011, Kursa_PRB2011, Villois_PRF2017}. This nontrivial configuration is well-captured by an MPS with $\chi_{\text{max}} = 80$, corresponding to a memory of $0.002\%$ with respect to the DNS. 

\begin{figure}[t]
    \centering
    \includegraphics[width= \linewidth,trim={5.0cm 0.0cm 0.0cm 4.0cm},clip]{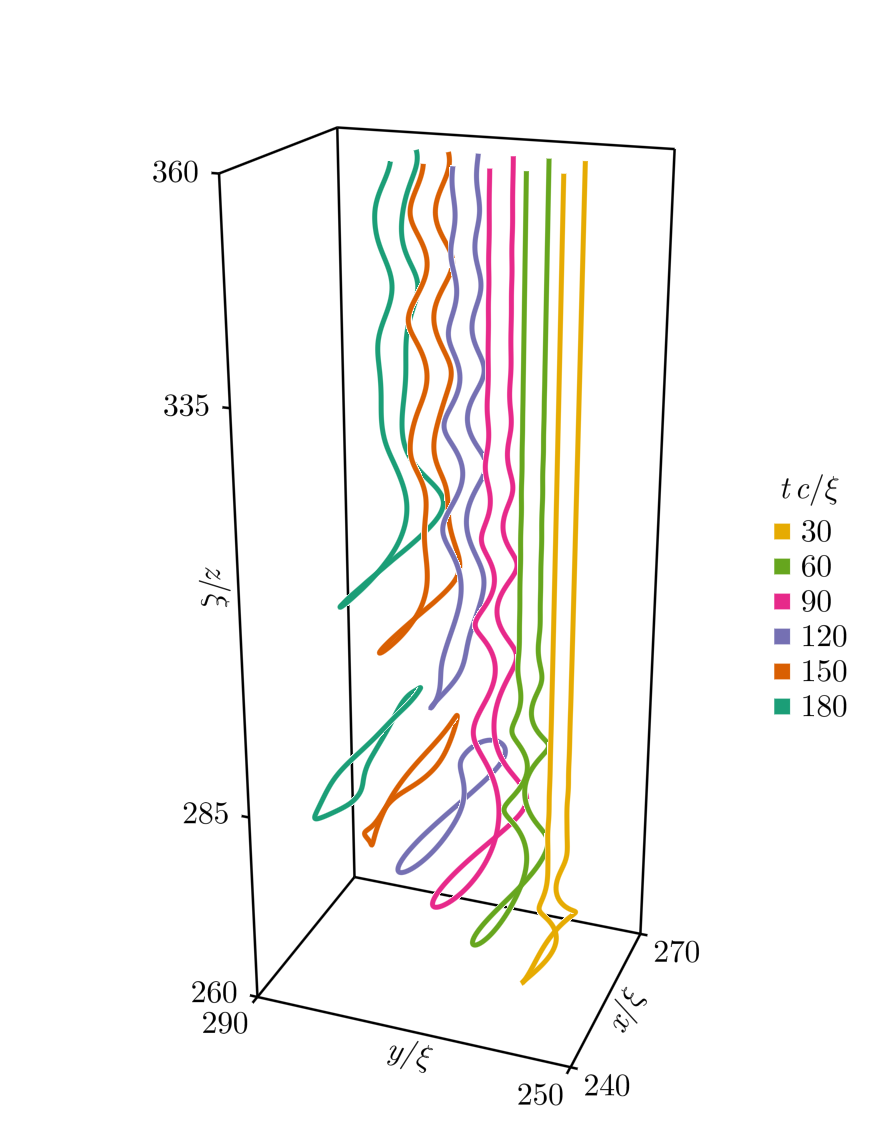}
    \caption{Generation of Kelvin waves after the reconnection shown in Fig.~\ref{fig:VortexReconnectionInit} for the vortex dipole along the $z$-axis. We show the vortex lines with a resolution of $\zeta = \xi/4$ at different times $t$ inside the box $[240\,\xi, 270\,\xi]\times[250\,\xi, 290\,\xi]\times[260\,\xi, 360\,\xi]$ with $\chi_{\text{max}} = 260$.}
    \label{fig:VortexReconnectionKelvin}
\end{figure}

The subsequent dynamics associated with the reconnected U-shaped vortex lines aligned along the vertical $z$-axis is shown in Fig.~\ref{fig:VortexReconnectionKelvin}. 
This plot exhibits Kelvin waves, which are helical oscillations of the vortex lines~\cite{Barenghi_2016}, induced by reconnection as they propagate upwards. The close-up region from Fig.~\ref{fig:VortexReconnectionKelvin} requires a system size $L \gtrapprox 200\, \xi$. In contrast, the initial vortex reconnection could in principle be studied with a smaller computational domain of $L \gtrapprox 40 \, \xi$. As the Kelvin waves propagate along the vortex lines, the nodes of the oscillation join and produce a second reconnection from which a vortex ring emanates. The complexity of the physics to be resolved increases, hence the necessary bond dimension also has to grow. Therefore, we fix $\chi_{\text{max}} = 260$, resulting in a memory which saturates around $0.03\%$ of that of DNS. With this level of compression, that is four orders of magnitude lower than the DNS memory size, we only require $\sim 40 \, \text{MB}$ to store the results at each timestep. 
 
\begin{figure*}[ht]
%\begin{figure*}[b!]
    \centering
    \includegraphics[width=\linewidth,trim={0.0cm 0.0cm 0.0cm 0.0cm},clip]{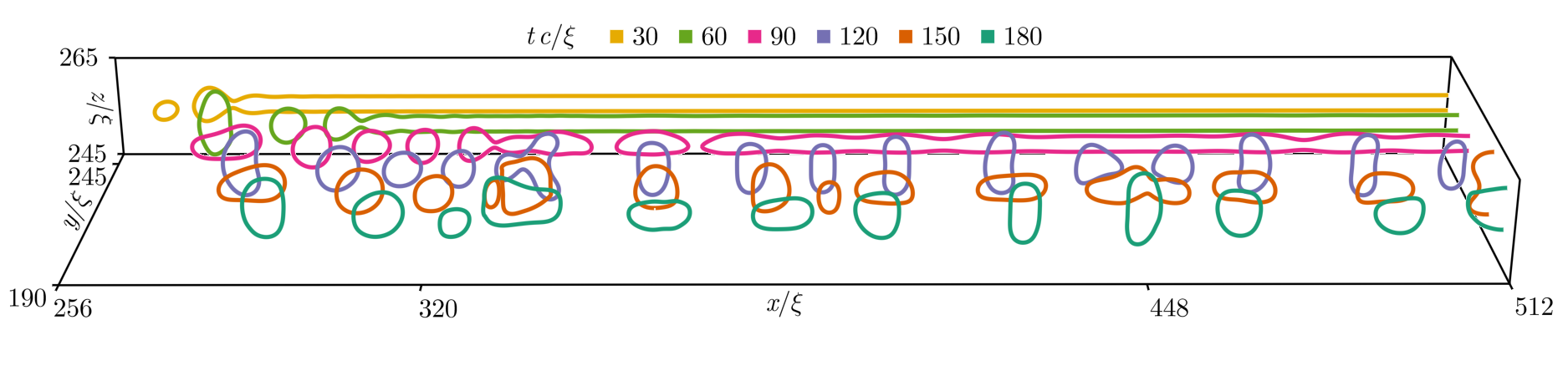}
    \vspace{-1.2cm}
    \caption{Vortex line collapse due to perturbations from the reconnection shown in Fig.~\ref{fig:VortexReconnectionInit} for the vortex dipole aligned along the $x$-axis. We show vortex lines with a resolution $\zeta = \xi/4$ for different times $t$ inside a box $[256\,\xi, 512\,\xi]\times[190\,\xi, 250\,\xi]\times[245\,\xi, 265\,\xi]$ with $\chi_{\text{max}} = 260$.}
    \label{fig:VortexReconnectionBreaking}
\end{figure*}

\begin{figure}[b]
    \centering
    \includegraphics[width=\linewidth]{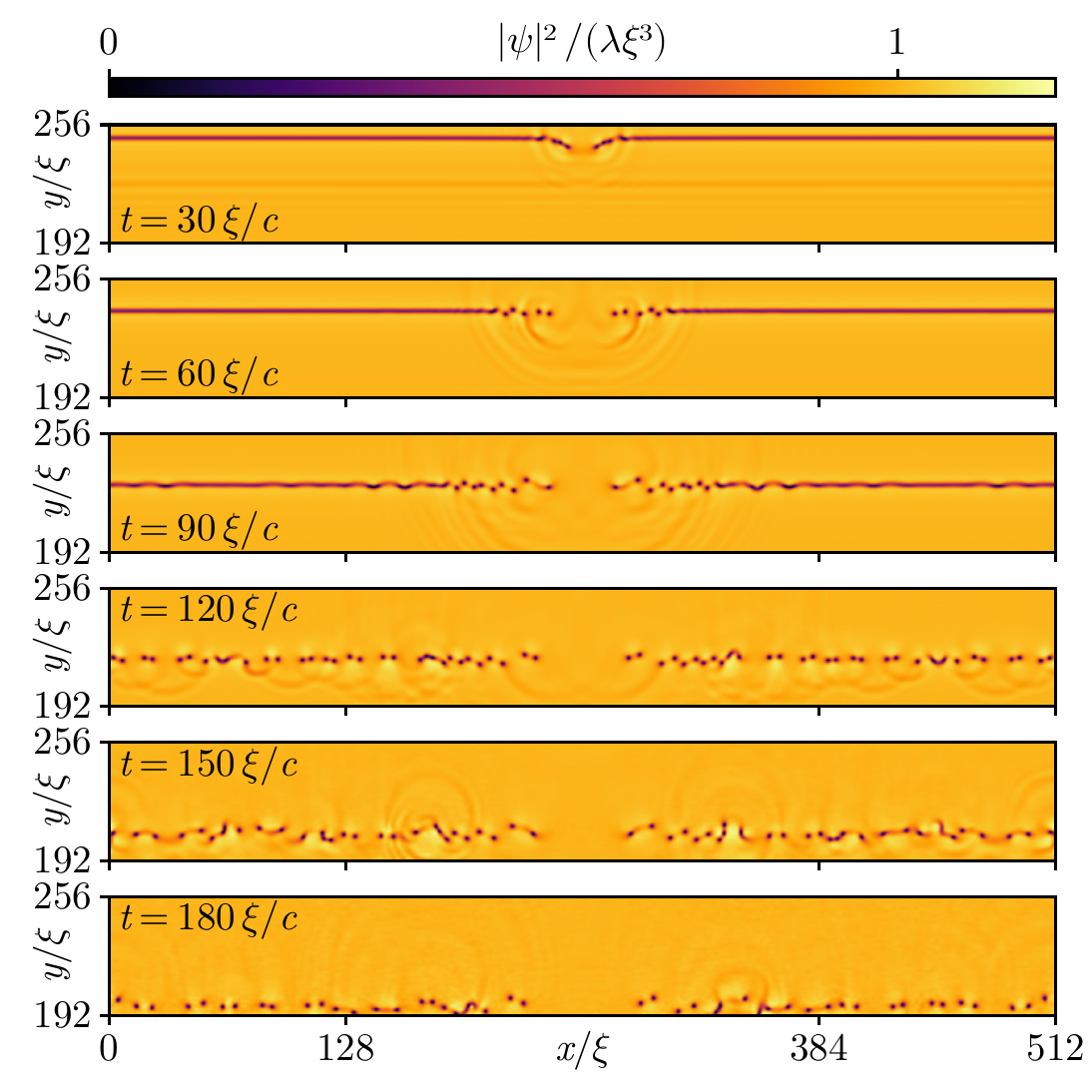}
    \caption{Density profile $|\psi|^2$ cut at $z=L/2$ and different times $t$ for the vortex dipole along the $x$-axis shown in Fig.~\ref{fig:VortexReconnectionBreaking}. The low-density horizontal lines indicate the vortex lines. Each pair of vortices (singularities) indicates the generation of a vortex ring.}
    \label{fig:VortexReconnectionWaves}
\end{figure}

A more striking phenomenon arises with the pair of vortex lines aligned along the $x$-axis. We show this in Fig.~\ref{fig:VortexReconnectionBreaking} at different times after the first reconnection. Here, the generation of vortex rings via the propagation of Kelvin waves occurs more rapidly than in Fig.~\ref{fig:VortexReconnectionKelvin} because of the shorter intervortex separation ($d = 4 \, \xi$). The initial stage of the vortex ring cascade has been discussed using the GP equation in Ref.~\cite{Kerr_PRL2011} and the vortex filament model in Ref.~\cite{Kursa_PRB2011}. Through our simulation, we observe that the rate at which the vortex ring cascade occurs varies with longer evolution times. To illustrate the different propagation velocities, we show in Fig.~\ref{fig:VortexReconnectionWaves} the cut of the density profile at $z = L/2$. It is computationally economical to build projections to a plane in the MPS ansatz, such as the $xy$ shown in Fig.~\ref{fig:VortexReconnectionWaves}, where all length scales of the unplotted axis, $z$ in the example, are contracted. This plot indicates that for $t \lesssim 90 \, \xi/c$ the production of vortex rings (identified by pairs of singularities in the density cut) coincides with the passage of the sound wave emitted by the reconnection. This comparison implies that the ring generation occurs at a rate equal to the speed of sound. However, between $t = 90 \, \xi/c$ and $120 \, \xi/c$, the small perturbations associated with the Kelvin wave propagate faster than the sound wave produced by the initial reconnection. These perturbations to the vortex line correspond to a Crow instability~\cite{Berloff_JPA2001}, which collapses the vortex dipole into a state made up of multiple vortex rings that coexist within the bath of radiated sound waves.

\section{Quantum turbulence}\label{sec:QuantumTurbulence}

Having benchmarked the quantics MPS time-evolution method in different dimensions, we now discuss its application to more complex dynamical simulations. Here, multiple nonlinear excitations coexist in the system, and their interactions give rise to emergent phenomena that have attracted significant interest in the quantum turbulence field. In 1D, we study the case of the soliton gas, a state in which multiple dark solitons govern the dynamical properties of integrable turbulence~\cite{Zakharov_SAM2009, Suret_PRE2024}. The soliton gas has been the focus of much interest in diverse fields, including optics~\cite{Suret_PRS2023}, liquid crystals~\cite{Pismen_1999, Aranson_RMP2002}, and water waves~\cite{Suret_PRL2020}. These physical systems are modeled by the nonlinear Schrödinger equation, for which there is a close correspondence to the GP. We then consider 2D and 3D quantum turbulence, characterized by the presence of multiple vortices~\cite{Barenghi_2016, Barenghi_2023}. 

In Sec.~\ref{sec:TurbulenceGeneration}, we generate different turbulence states containing multiple nonlinear excitations using DNS and focus on how the bond dimensions of its MPS encoding, for a fixed infidelity criteria, scale with the system size $L$ and number of excitations. This analysis serves as a basis for understanding the truncation requirements of turbulent states represented by MPS and comparing the memory with that used by DNS. Afterwards, in Sec.~\ref{sec:TurbulenceDynamics}, we investigate the dynamics of the initial turbulent states using the quantics MPS time-evolution method for specific examples in each dimension and compare with the results found in Sec.~\ref{sec:TurbulenceGeneration}.

\subsection{Generation of turbulence states} \label{sec:TurbulenceGeneration}

\begin{figure*}[ht]
    \centering
    \includegraphics[width=\linewidth]{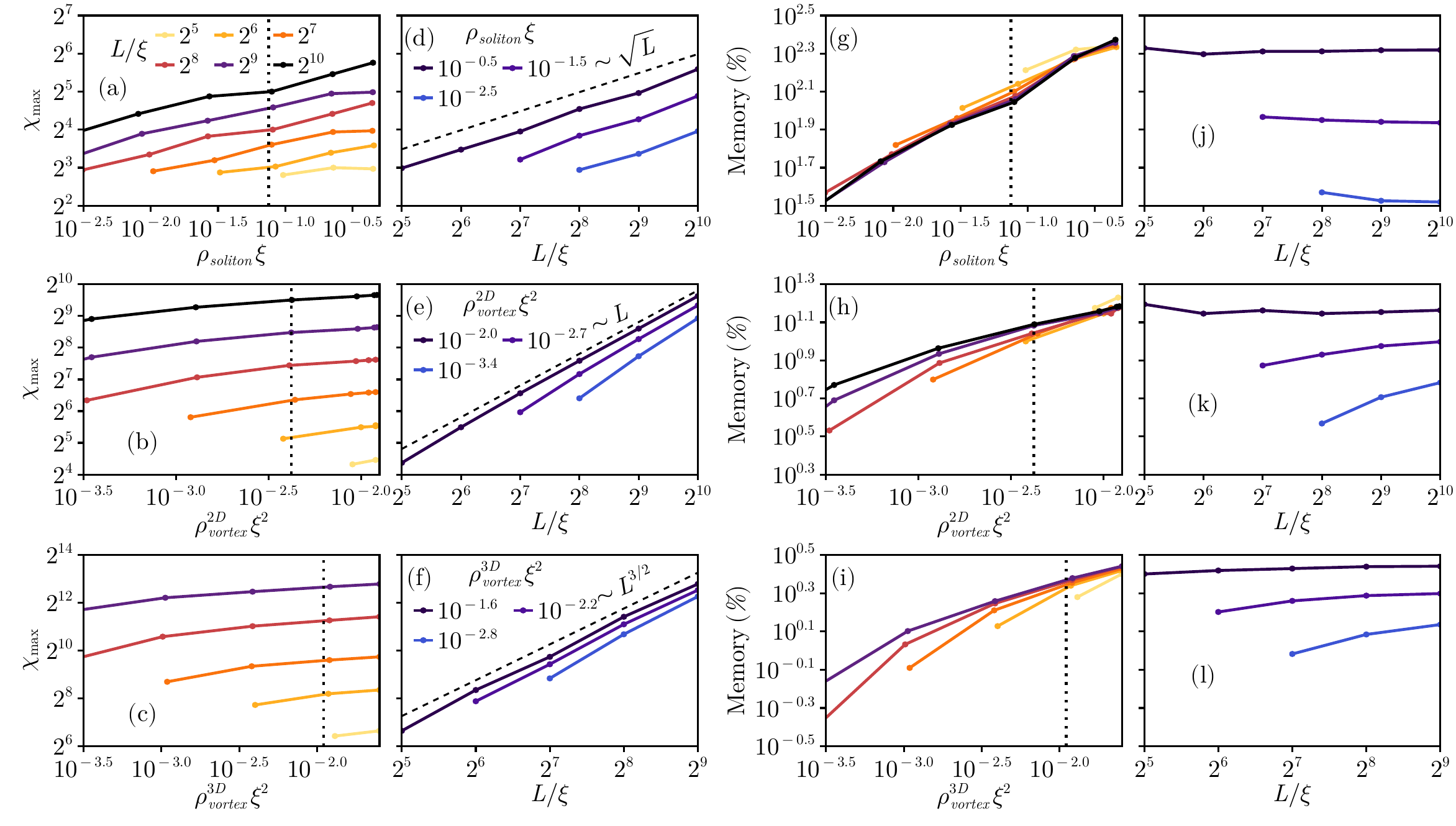}
    \caption{
        Truncation parameters for the MPS representation of initial turbulent states created using DNS with an infidelity of $I \approx 10^{-6}$. (a-f) Maximum bond dimension ($\chi_{\text{max}}$) and (g-l) memory percentage with respect to DNS as a function of density $\rho_{\text{Soliton}}$ or $\rho_{\text{Vortex}}^{2D/3D}$ (a-c, g-i) and system size $L$ (d-f, j-l). Each point in (a-c, g-i) corresponds to six averaged random states, and linear interpolations of these results are used to obtain the data in (d-f, j-l). We fix a resolution of $\xi/4$. Since for 3D, the latter corresponds to extensive grids, we extrapolate the results from coarser resolutions of $\xi/2$ and $\xi$. The dotted vertical lines correspond to the test densities in Sections \ref{sec:SolitonGas} (a, g), \ref{sec:PointVortexDistribution} (b, h), and \ref{sec:VortexTangle} (c, i). The dashed lines indicate power laws of $\sim \sqrt{L}$ (d), $\sim L$ (e), and $\sim L^{3/2}$ (f).
    }
    \label{fig:Scaling}
\end{figure*}

We start with the preparation of the turbulence states using DNS. For this generation, random values are assigned to the wave function in an initial coarse grid, and a linear interpolation is applied to reconstruct the field at the desired grid resolution. This process generates multiple phase jumps across the system, and a short imaginary time evolution transforms them into the corresponding excitations, either dark solitons in 1D or quantized vortices in 2D and 3D. This last step works particularly well for 2D and 3D since vortices are topologically protected and can only disappear by annihilation with each other during imaginary time evolution. For 1D, this is not the case as the solitons decay uniformly to the corresponding ground state. Hence, to generate the 1D states, we fix the phase of the wave function during the imaginary time evolution and change the interpolation from linear to cubic to obtain a smooth initial phase field. We call this the \emph{random phase interpolation} method, which is a modification of the proposed one in Ref.~\cite{Kobayashi_PRL2005}. Further details of the methodology are provided in Appendix~\ref{sec:RandomPhaseInterpolation}.

Using the method above, we generate the initial states for each dimension with different densities of excitations. In 1D, we quantify the amount of solitons using the following definition~\cite{Meng_NJP2023}
\begin{align} \label{eqn:SolitonDensity}
    \rho_{\text{Soliton}} = \frac{1}{L}\int_0^L  \frac{1-\frac{|\psi|^2}{\lambda \xi}}{2} \dd x \, .
\end{align}
The latter is equal to $1/L$ for a single static soliton and decreases to zero as the speed of the soliton approaches $c$ when $|\psi|^2 \approx \lambda \xi$. For 2D and 3D, we use the vortex counting method from Ref.~\cite{Phillips_PRE2015} based on the winding number around the vortex core. Here we define the vortex density $\rho_{\text{Vortex}}^{2D/3D}$ as the number of vortices (length of vortex lines) divided by the area (volume) of the box, respectively. Then, we build the quantics MPS representation of the random initial state using iterative SVDs, as shown in Fig.~\ref{fig:TensorNetworkExample}(c). For this, $\varepsilon$ is varied until a desired infidelity of $I \approx 10^{-6}$ with the original state is reached. We use the staircase orderings for 2D and 3D, and choose the $xyz$ axis order for the latter. Given that correlations are equally distributed among all directions for a random excitation configuration, choosing another axis order does not appreciably change the results.

We show the resulting maximum bond dimensions $\chi_{\text{max}}$ in Fig.~\ref{fig:Scaling}. In all cases, we find that $\chi_{\text{max}}$ grows with the densities $\rho_{\text{Soliton}}$ or $\rho_{\text{Vortex}}^{2D/3D}$ for a fixed $L$ (Figs.~\ref{fig:Scaling}(a-c)) as well as with the system size $L$ for a constant density (Figs.~\ref{fig:Scaling}(d-f)). This means that $\chi_{\text{max}}$ grows with the number of excitations, whether these are solitons or vortices. In particular, by fixing the densities for $d$ spatial dimensions, we find a power law scaling of $\chi_{\text{max}} \sim \sqrt{L^d}$, which is denoted by the dashed lines in Figs.~\ref{fig:Scaling}(d-f). We know from the iterative SVD construction of an MPS (Fig.~\ref{fig:TensorNetworkExample}(c)) that its largest bond dimension without truncating singular values is $\chi_{\text{max}}^{\text{SVD}} = D^{N/2}$. Since the total number of grid points is $M = D^N$ we have $\chi_{\text{max}}^{\text{SVD}} = \sqrt{M}$ in the untruncated limit. For the results of Fig.~\ref{fig:Scaling}(d-f), since we fix the spatial resolution, we have $M \sim L^d$ grid points and obtain $\chi_{\text{max}} \sim \sqrt{M}$, a similar behavior as $\chi_{\text{max}}^{\text{SVD}}$. Here, the constant of proportionality for $\chi_{\text{max}}$ is in general much lower than one since we find considerable memory compression.

Although $\chi_{\text{max}}$ grows with $L$, the corresponding memory that is required to store the MPS scales differently, given the nonuniform distribution of bond dimensions. To illustrate this, in Fig.~\ref{fig:Scaling} we also plot the percentage of memory with respect to that required by DNS. Strikingly, the memory percentage is solely a function of the corresponding density (Figs.~\ref{fig:Scaling}(g-i)). We emphasize this by fixing a density and observing a converging behavior to a fixed percentage when increasing $L$ (Figs.~\ref{fig:Scaling}(j-l)). Furthermore, we note that the memory percentage curve decreases by around a factor of 10 when increasing the dimensionality of the system, a behavior observed across the various examples considered in Sections \ref{sec:DarkSoliton}, \ref{sec:VortexDipole}, and \ref{sec:VortexRing}.

These results emphasize that the main computational saving of the method is expected to occur for 3D simulations. According to Fig.~\ref{fig:Scaling}(i), the memory percentage is capped at $10^{0.5} \% \approx 3\%$ of that which DNS requires in the system of highest density. The vortex line density range considered here includes values such as $\rho_{\text{Vortex}}^{3D} \sim 10^{-3} \, \xi^{-2}$, which is large enough to feature complex vortex line dynamics~\cite{Stagg_PRA2016}. Furthermore, for the low-density examples such as the vortex line reconnection from Sec.~\ref{sec:VortexLineReconnection}, we achieve even lower memory percentages of $0.03\%$. As a comparison, the largest system size used in the literature for simulating vortex reconnections corresponds, to the best of our knowledge, to a grid of $5760^3 \sim 10^{11}$ points. This calculation requires a vast memory of $2848 \, \text{GB}$ and was performed with a qubit lattice algorithm, highly parallelized with $12288$ cores~\cite{Yepez_PRL2009}. This large-scale simulation uses the same vortex line density as the case studied through Sec.~\ref{sec:VortexLineReconnection} ($\rho_{\text{Vortex}}^{3D} = 4L/L^3 = 4L^{-2} = 2 \times 10^{-5} \, \xi^{-2}$). Hence, assuming the same memory percentage of $0.03\%$ of that required by DNS, we would need only $\sim 875\, \text{MB}$ with our MPS method. Remarkably, our GPU with memory of $40 \, \text{GB}$ is more than enough to tackle this job successfully.

\subsection{Turbulence dynamics}\label{sec:TurbulenceDynamics}

In the following, we analyze the dynamics of the states obtained in Sec.~\ref{sec:TurbulenceGeneration} with a specific density for each dimension, indicated by dotted lines in panels (a-c) and (g-i). We focus on how the quantics MPS time evolution captures the dynamics driven by the damped GP equation. For 1D (Sec.~\ref{sec:SolitonGas}), we use the initial quantics MPS obtained with iterative SVDs after using the random interpolation method since the grid sizes for this low dimension are not prohibitively large. On the contrary, in 2D (Sec.~\ref{sec:PointVortexDistribution}) and 3D (Sec.~\ref{sec:VortexTangle}), using SVDs can be computationally expensive and requires substantial amounts of memory for a large number of grid points ($\sim 2^{30}$, corresponding to $\sim 30$ qubits). In these cases, we can directly mimic the random phase interpolation method in the quantics MPS format. This requires starting with a random MPS for the coarser grid and using linear interpolation operators represented by tensor networks~\cite{Gourianov_2022}, which resolves the remaining length scales with the introduction of additional qubits. This extension to the MPS framework is detailed in Appendix~\ref{sec:RandomPhaseInterpolation}.

\subsubsection{1D Soliton gas} \label{sec:SolitonGas} 

\begin{figure}[t]
    \centering
    \includegraphics[width=\linewidth]{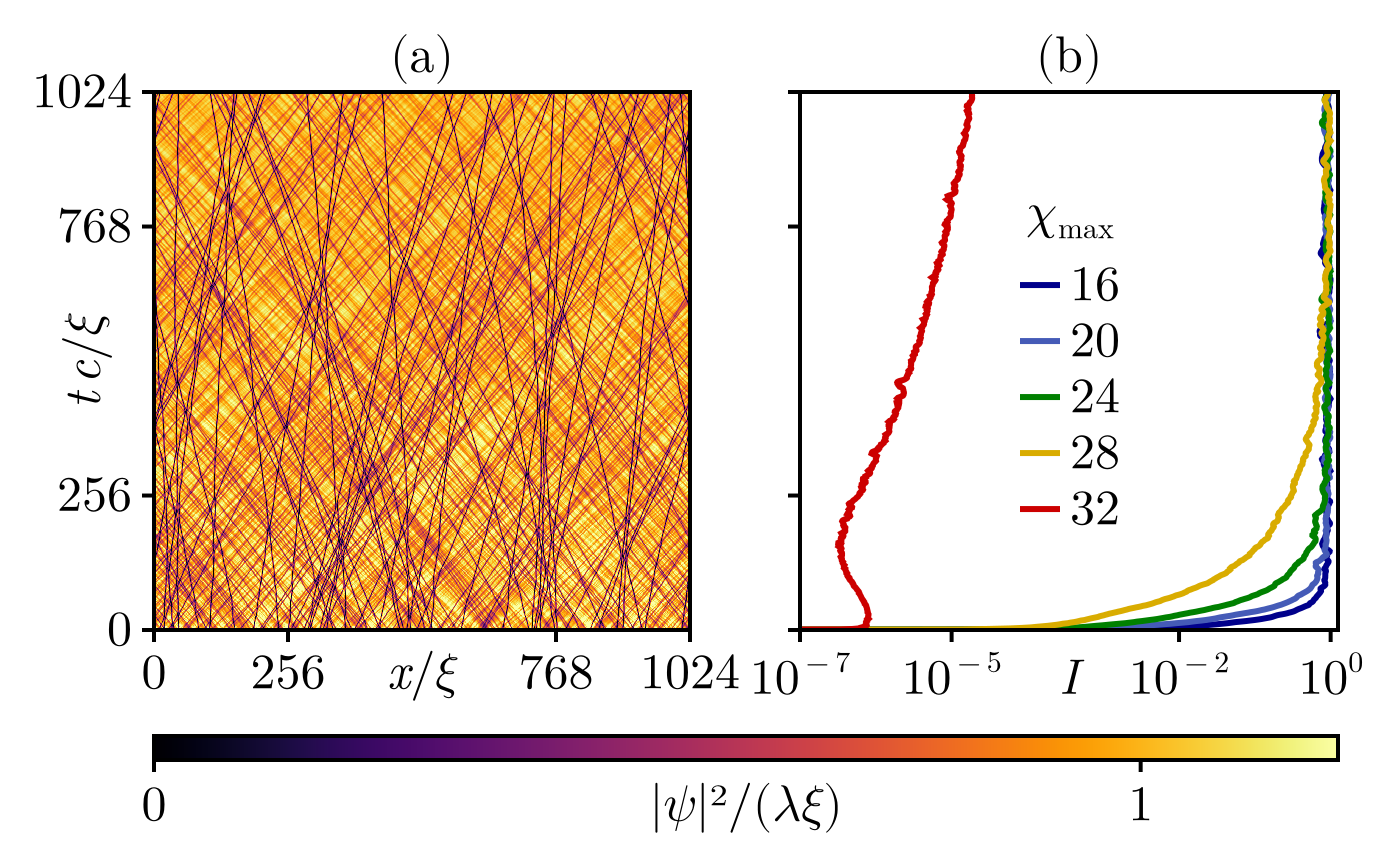}
    \caption{
        Evolution of soliton gas with initial density $\rho_{\text{Soliton}} = 7.5 \times 10^{-2} \, \xi^{-1}$ in a domain of size $L = 1024\,\xi$ with a resolution of $\xi/4$ and $\Delta t = 2^{-6} \, \xi/c$. (a) Density profile $|\psi|^2$ for $\chi_{\text{max}} = 32$. (Right) Infidelity $I$ with respect to the DNS for different $\chi_{\text{max}}$ values.
    }
    \label{fig:SolitonGas_Evolution}
\end{figure}

Here we consider a soliton density of $\rho_{\text{Soliton}} = 7.5 \times 10^{-2} \, \xi^{-1}$ in a box of size $L = 1024 \, \xi$ and observe its dynamics until $t = L/c$. We also set a damping coefficient of $\gamma = 10^{-3}$ to suppress the sound waves in the system without annihilating the dark solitons. In Fig.~\ref{fig:SolitonGas_Evolution} we plot the evolution of the density profile for $\chi_{\text{max}} = 32$ (left panel), which has an infidelity of $I \lessapprox 10^{-5}$ (right panel). 
Thus, for this value of the bond dimension, the MPS calculation accurately captures the complex dynamics of multiple soliton trajectories (represented by dark lines) and their collisions. For lower bond dimensions, the trajectories differ appreciably, as indicated by the corresponding infidelity curves that rapidly approach 1. This result is in agreement with the bond dimension obtained in Fig.~\ref{fig:Scaling}(a) (dotted line). Due to the low dimensionality of the problem, there is a limited amount of compression in the memory with respect to the DNS (dotted line in Fig.~\ref{fig:Scaling}(g)).

\subsubsection{2D Point-vortex distribution} \label{sec:PointVortexDistribution}

\begin{figure}[t]
    \centering
    \includegraphics[width=\linewidth]{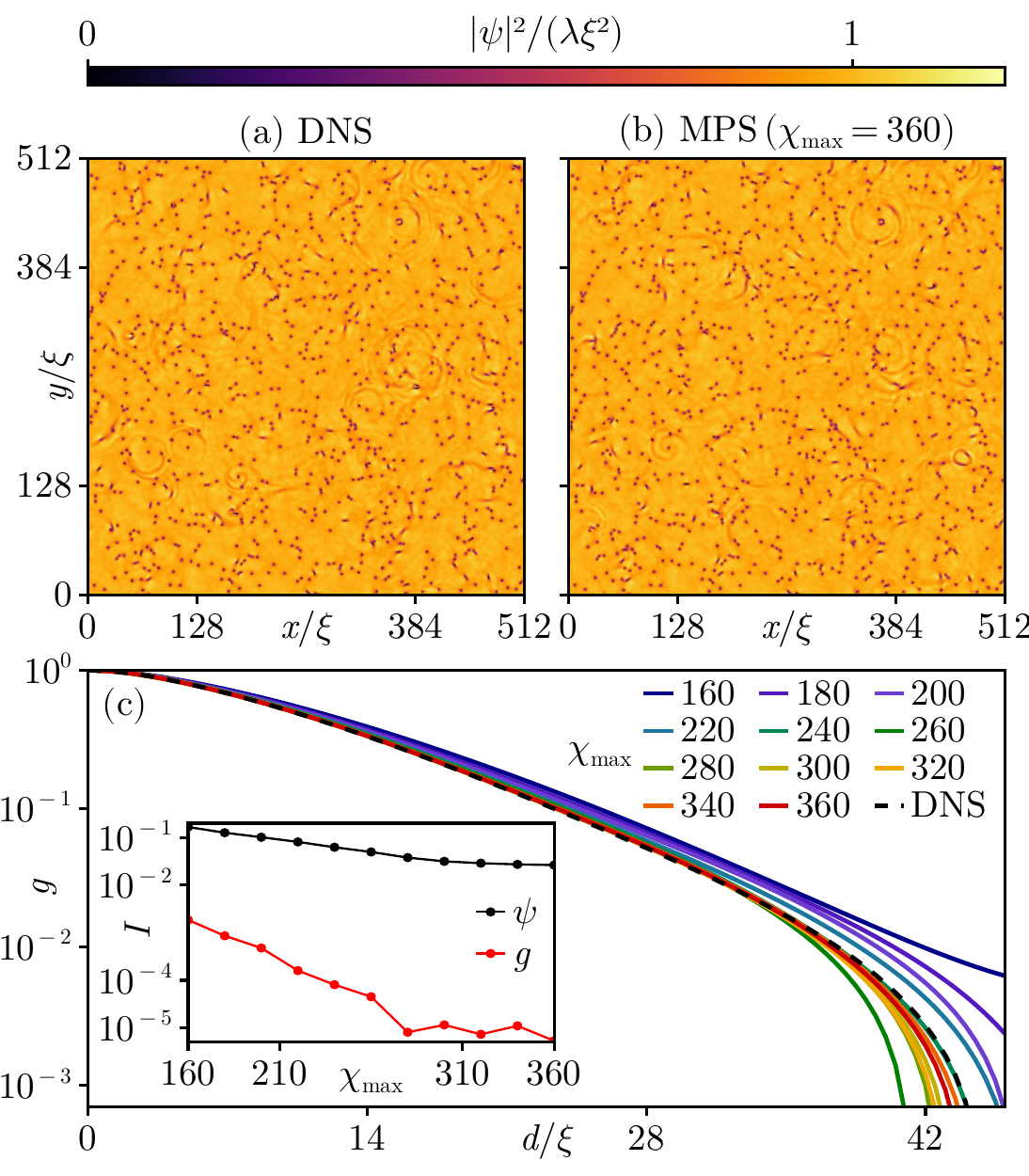}
    \caption{
        Density profiles of the DNS (a) and MPS simulation with $\chi_{\text{max}} = 360$ (b) for the turbulent state evolved up to $t = 200\, \xi/c$. (c) Correlation function $g$ for various $\chi_{\text{max}}$ values with the MPS method and the DNS. The inset in (c) shows the infidelity with respect to the DNS of the wave function $\psi$ and correlation function $g$ as a function of $\chi_{\text{max}}$. The system size is $L = 512 \, \xi$ with a resolution of $\xi/4$ and $1100$ initial vortices. The evolution is performed with a damping of $\gamma = 10^{-2}$ using $\Delta t = 2^{-6} \, \xi/c$.
    }
    \label{fig:QT2D_Profiles}
\end{figure}

We choose a system size of $L = 512 \, \xi$ with a vortex density of $\rho_{\text{Vortex}}^{2D} = 4.2 \times 10^{-3} \, \xi^{-2}$. We let the state evolve for $t = 200\, \xi/c$ with $\gamma = 10^{-2}$. The density profiles of the final states are shown in Fig.~\ref{fig:QT2D_Profiles}, for DNS (panel (a)) and MPS with $\chi_{\text{max}} = 360$ (panel (b)). Here, we observe similar vortex distributions with small discrepancies.

Given that these simulations result in chaotic behavior, a point-to-point comparison is not adequate to evaluate the accuracy of the MPS evolution. Instead, we determine the extent to which the MPS evolution captures spatial-averaged correlations across the system. We consider the two-point correlation function $g$, given by
\begin{align} \label{eqn:CorrelationFunction}
    g(l) = \frac{1}{ 2\pi \int |\psi|^2 \dd^2 \br}\oint \int \psi^*(\vb{r}) \psi(\vb{r}-l\hat{n}) \dd^2 \br \dd \theta \, ,
\end{align}
where $\hat{n} = (\cos \theta, \sin \theta)$\footnote{The function $g$ is real-valued for periodic boundary conditions since the imaginary parts cancel out after the angular integration.}, used to characterize first-order correlations in condensates~\cite{Pethick_2008}, optics~\cite{Mandel_1995}, and condensed matter~\cite{Mahan_2000} systems. We plot $g$ for different $\chi_{\text{max}}$ values in Fig.~\ref{fig:QT2D_Profiles}(c). The correlation functions that match the DNS result correspond to $\chi_{\text{max}} \geq 280$. For these bond dimensions, the largest discrepancies occur when $g \leq 10^{-2}$, mainly due to the small deviations of the vortex positions. To quantify the difference in correlations with respect to the DNS results, we use the infidelity measure applied to $g$ as
\begin{align}
    I(g_1, g_2) = 1 - \frac{(g_1 \cdot g_2)}{\sqrt{(g_1 \cdot g_1) \, (g_2 \cdot g_2})}.
\end{align}
Here $g_1$ and $g_2$ are the correlation functions of the compared states and $(g_1 \cdot g_2) = \int g_1(x) g_2(x) \dd x$ over the corresponding spatial domain. In the inset of Fig.~\ref{fig:QT2D_Profiles}(c) we find infidelities of $I \approx 10^{-5}$ for $g$ (red curve) and $\chi_{\text{max}} \geq 280$. On the other hand, the infidelity for the wave function (black curve) is larger than $I \geq 10^{-2}$. Hence, even though the DNS and MPS final states differ due to the chaotic nature of the turbulent dynamics, the statistical properties of the turbulence are well reproduced. This representation is achieved for low-enough bond dimensions such that the memory usage reaches $22\%$ (for $\chi_{\text{max}} = 280$) compared to that which DNS requires.

\begin{figure}[b]
    \centering
    \includegraphics[width=\linewidth]{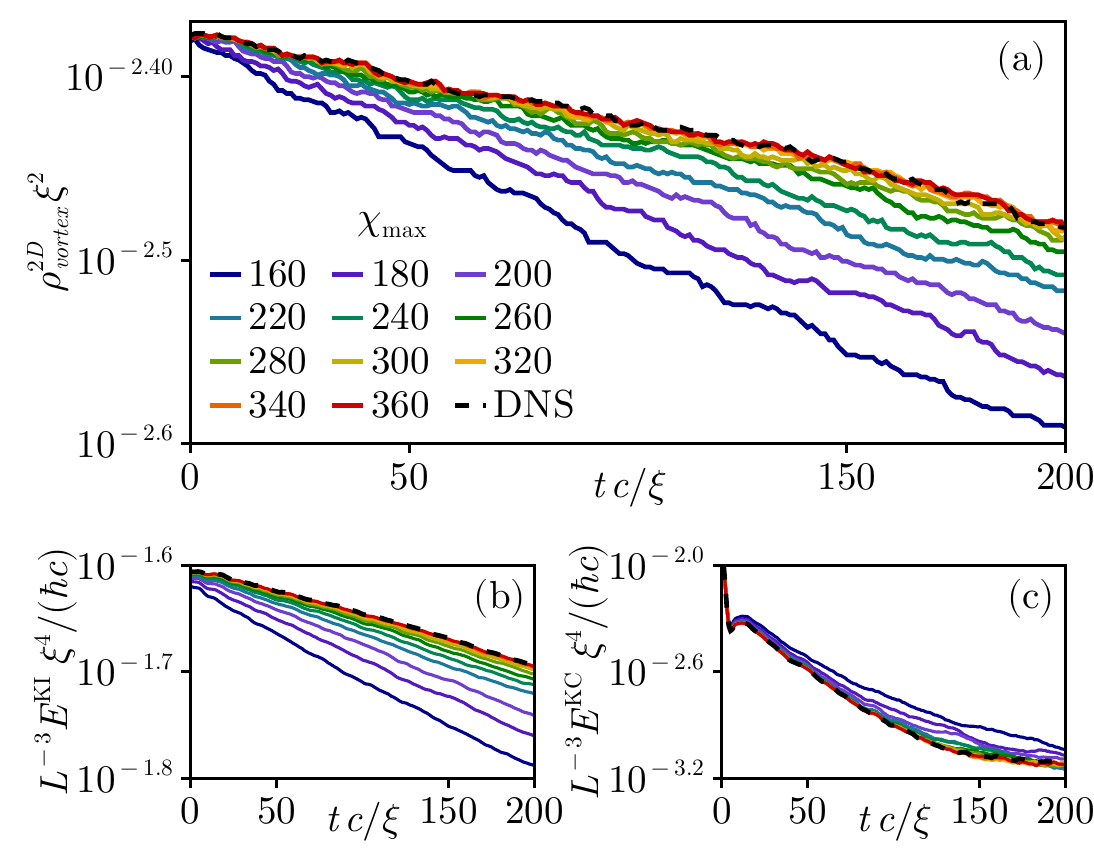}
    \caption{(a) Vortex density $\rho_{\text{Vortex}}^{2D}$ as a function of time $t$ for the simulation of Fig.~\ref{fig:QT2D_Profiles}, obtained with DNS and MPS with various bond dimensions $\chi_{\text{max}}$. Kinetic energy decomposition in its incompressible $E^{\text{KI}}$ (b) and compressible $E^{\text{KC}}$ (c) contributions as a function of time.}
    \label{fig:QT2D_VortexNumber}
\end{figure}

In addition to the vortex distribution, we are also interested in how other averaged quantities evolve with time using the MPS representation. In Fig.~\ref{fig:QT2D_VortexNumber} we plot the vortex density $\rho_{\text{Vortex}}^{2D}$ as a function of time. 
We observe a decay of $\rho_{\text{Vortex}}^{2D}$ for the DNS (black dashed line), evidence of vortex pair annihilations for $\gamma > 0$ as shown in Fig.~\ref{fig:VortexDipoleEvolution}(g-i) and collisions between vortex dipoles~\cite{Baggaley_PRA2018}. For low $\chi_{\text{max}}$ values, artificial damping results in the rapid annihilation of vortex pairs. However, we find that for bond dimensions larger than  $\chi_{\text{max}} \approx 320$, the vortex density decays at the same rate as that seen in the DNS, thus suppressing the artificial damping observed for lower values of $\chi_{\text{max}}$. This bond dimension corresponds to a memory of $27\%$ with respect to the DNS.

The expected decay of the vortex number, which occurs under the dynamics of the damped GP equation evolution, is also evident in the time dependence of the kinetic energy. To observe this, we first factorize the kinetic energy into its incompressible and compressible contributions~\cite{Nore_PF1997, Numasato_PRA2010, Nowak_PRA2012}:
\begin{align}
    E^{\text{KI}} = \int \frac{m}{2} |\vb{w}^{\text{I}}|^2 \dd^2 \br \, , \quad
    E^{\text{KC}} = \int \frac{m}{2} |\vb{w}^{\text{C}}|^2 \dd^2 \br. \label{eqn:Kinetic}
\end{align}
The compressible component is associated with the sound waves in the system, whereas the incompressible component is attributed to the quantized vortices. To define the energy decomposition, we use the Mandelung transformation $\psi = \sqrt{\rho} e^{i\theta}$ to obtain the density $\rho$ and velocity $\vb{v} = \frac{\hbar}{m} \vb{\nabla} \theta$~\cite{Pitaevskii_2016, Pethick_2008, Barenghi_2016}. We can then define the pseudo-velocity $\vb{w} = \sqrt{\rho}\vb{v}$ and use the Helmholtz decomposition to recover the incompressible $\vb{w}^{\text{I}}$ and compressible $\vb{w}^{\text{C}}$ components. In Fig.~\ref{fig:QT2D_VortexNumber}(b,c), we show each contribution as a function of time.
We note that the incompressible kinetic energy (panel (b)) exhibits a trend that is analogous to the time-dependence of the vortex number shown in Fig.~\ref{fig:QT2D_VortexNumber}(a). This result is expected given that $E^{\text{KI}}$ corresponds to the energy associated with the vortex excitations. The more rapid vortex annihilation rate that occurs for low values of $\chi_{\text{max}}$ leads to an increased generation of sound waves, which increases the compressible kinetic energy (panel (c)). As seen before, to correctly recover the decay rates of the DNS, we require $\chi_{\text{max}} \geq 320$, in analogy to the values quoted for the time-dependence of the vortex number.

\begin{figure}[t]
    \centering
    \includegraphics[width=\linewidth]{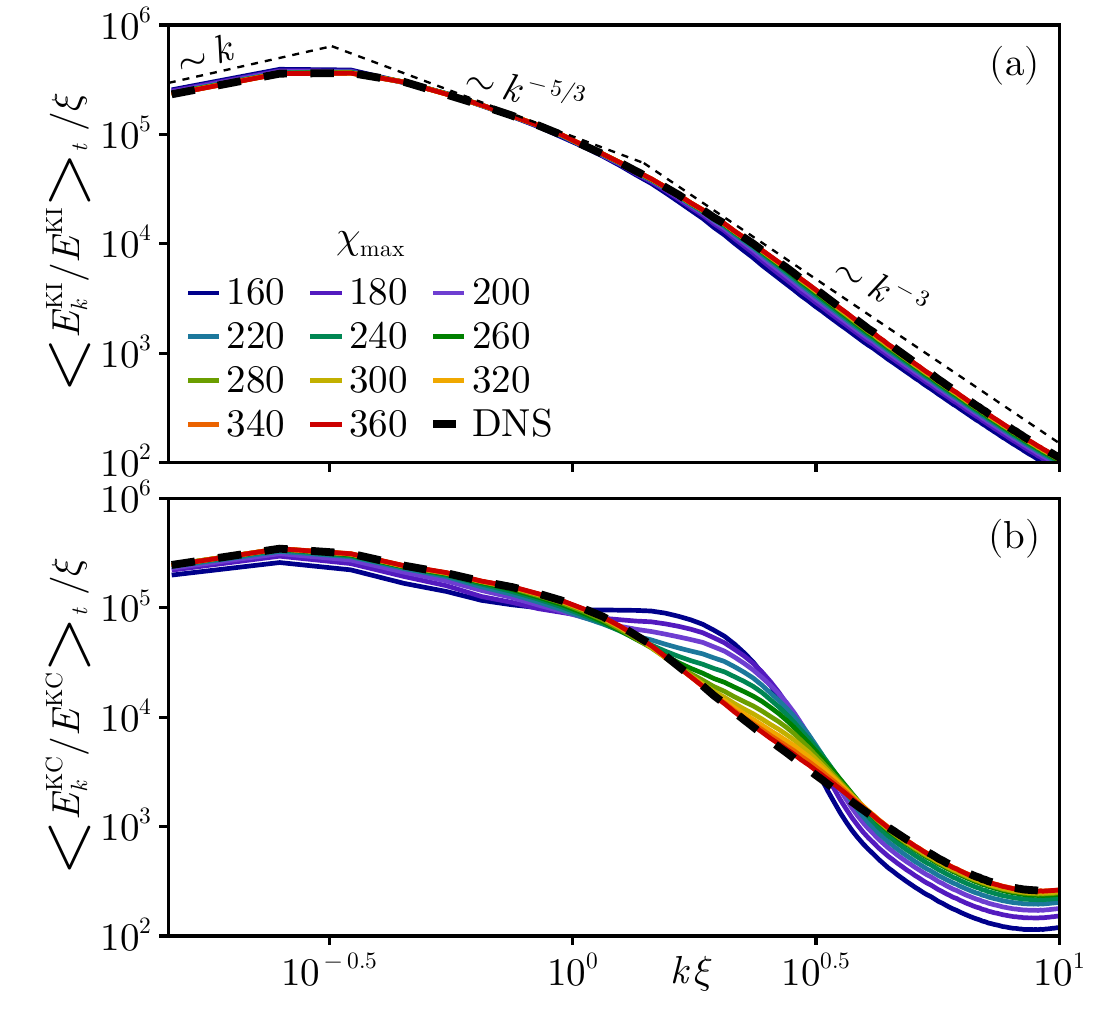}
    \caption{
        Time-averaged scaled kinetic energy spectrum for the incompressible $\left< \overline{E}^{\text{KI}}_{k} / E^{\text{KI}} \right>_t$ (a) and compressible $\left< \overline{E}^{\text{KC}}_{k} / E^{\text{KC}} \right>_t$ (b) components. We use the parameters from Fig.~\ref{fig:QT2D_Profiles} and compare the DNS and MPS results with multiple $\chi_{\text{max}}$ values. The dashed lines in (a) indicate power laws of $k$, $k^{-5/3}$, and $k^{-3}$, from left to right.
    }
    \label{fig:QT2D_EnergiesSpectrum}
\end{figure}

One of the most important quantities that is often analyzed in the field of quantum turbulence is the kinetic energy spectrum, in analogy with Kolmogorov's analysis for classical fluids~\cite{Kraichnan_PF1967, Batchelor_PF1969, Barenghi_2016}. To assess how the time evolution of the MPS reproduces the energy cascade, we define the incompressible and compressible kinetic energy spectra as
\begin{align}
    \overline{E}^{\text{KI/KC}}_{k} = \oint %_{\hat{n}} 
    \frac{m}{2}|\vb{\overline{w}}^{\text{I/C}}_{\vb{k}}|^2 \dd \theta. \label{eqn:Spectra}
\end{align}
Here we use the Fourier transforms $\vb{\overline{w}}^{\text{I/C}}_{\vb{k}} = \mathcal{F} \left[\vb{w}^{\text{I/C}}\right]_{\vb{k}}$ and average over the direction of $\vb{k}$ with $\theta$. We show each spectrum in Fig.~\ref{fig:QT2D_EnergiesSpectrum} as $\left< \overline{E}^{\text{KI/KC}}_{k} / E^{\text{KI/KC}} \right>_t$ where we scale by the corresponding total kinetic energy $E^{\text{KI/KC}}$ and average over the time interval $t \in [0,200\, \xi/c]$. 
In Fig.~\ref{fig:QT2D_EnergiesSpectrum}(a), we observe an incompressible energy spectrum that is essentially independent of the value of $\chi_{\text{max}}$. The results are in excellent agreement with the spectrum recovered from the DNS. To further illustrate the efficacy of the MPS representation, we also compare the incompressible spectrum with the expected scaling behaviors reported at different length scale regions~\cite{Nowak_PRA2012, Bradley_PRX2012}. For large length scales, corresponding to small values of $k$, we observe a linear ($\sim k$) scaling that is independent of the vortex distribution. For small length scales, corresponding to large $k$ values, we capture the characteristic $\sim k^{-3}$ scaling due to the vortex core structure. In the intermediate region, we observe a spectrum which corresponds to the expected $k^{-5/3}$ scaling. Even though an insufficient bond dimension gives rise to spurious damping, the resulting vortex distribution preserves its main spectral properties. In fact, the main difference between the results obtained with large and small values of $\chi_{\text{max}}$ occurs for large $k$. We observe that, for the $\sim k^{-3}$ segment of the curve, the simulations performed with small values of $\chi_{\text{max}}$ fall off more rapidly. 

On the other hand, the generation of additional sound waves by unphysical vortex pair annihilation events for small values of $\chi_{\text{max}}$ drastically changes the compressible spectrum shown in Fig.~\ref{fig:QT2D_EnergiesSpectrum}(b). 
In particular, a ``shoulder'' emerges in the computed energy spectrum at values of $k$ that coincide with the length scale of the vortex core, leading to a bottleneck in the cascade. Hence, for the compressible spectrum, we recover the DNS results only for a sufficiently large value of $\chi_{\text{max}}$.

\begin{figure*}[t]
    \centering
    \includegraphics[width=\linewidth]{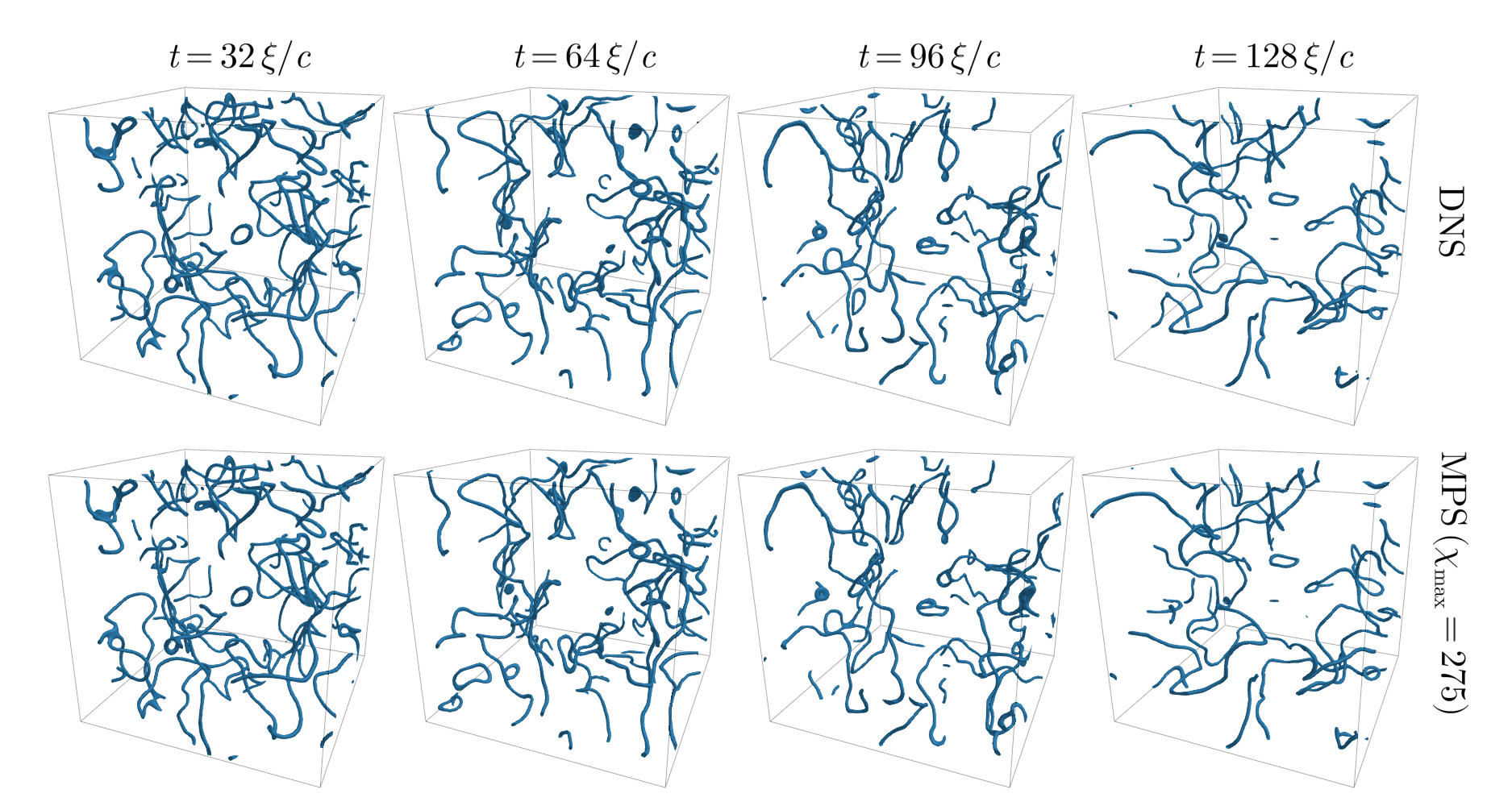}
    \caption{Vortex density profiles of a 3D quantum turbulent state for $L = 64\,\xi$,  $|\psi|^2 = 0.1 \, \lambda \xi^3$ and different times $t$ with $\Delta t = 2^{-6} \, \xi/c$. We show the DNS (upper row) and MPS (lower row) snapshots, with the latter truncated using $\chi_{\text{max}} = 275$.}
    \label{fig:QT3D_Profiles}
\end{figure*}

From the previous results, we conclude that to recover the principal characteristics of the 2D quantum turbulence evolution driven by the damped GP equation, a bond dimension of $\chi_{\text{max}} \approx 320$ is necessary. This value corresponds to a memory of $27\%$ of that required by DNS. The bond dimension is close to the expected value obtained from DNS data ($\chi_{\text{max}} = 357$, dotted line in Fig.~\ref{fig:Scaling}(e)), with a slight difference of $\sim 40$ retained singular values. The obtained memory of $27\%$, on the other hand, is above the corresponding result shown in Fig.~\ref{fig:Scaling}(h) of $12\%$ (dotted line). In this section, since we have varied $\chi_{\text{max}}$ rather than $\varepsilon$ as performed for Fig.~\ref{fig:Scaling}, less truncation is applied during the MPS evolution. This results in a larger memory usage compared to the DNS analysis shown in Fig.~\ref{fig:Scaling}(g).

\subsubsection{3D Vortex tangle} \label{sec:VortexTangle}

Finally, we simulate the dynamics of a 3D turbulent system using the MPS time evolution. We consider an initial state with a density of $\rho_{\text{Vortex}}^{3D} = 1.1 \times 10^{-2} \, \xi^{-2}$ in a domain with $L = 64\,\xi$ with the fixed resolution $\xi/4$. We evolve the turbulent state until $t = 128 \, \xi/c$ with $\gamma = 10^{-2}$. We show snapshot samples in Fig.~\ref{fig:QT3D_Profiles} for DNS (upper row) and TDVP (lower row), the latter with $\chi_{\text{max}} = 275$. We observe similar dynamics in both cases with small differences in the chaotic trajectories of the vortex lines, similarly to the results presented in Sec.~\ref{sec:PointVortexDistribution}. This example demonstrates the efficacy of the MPS ansatz: even for this turbulent state, we recover the dynamics using $7\%$ of the memory required by the DNS.

\begin{figure}[b]
    \centering
    \includegraphics[width=\linewidth]{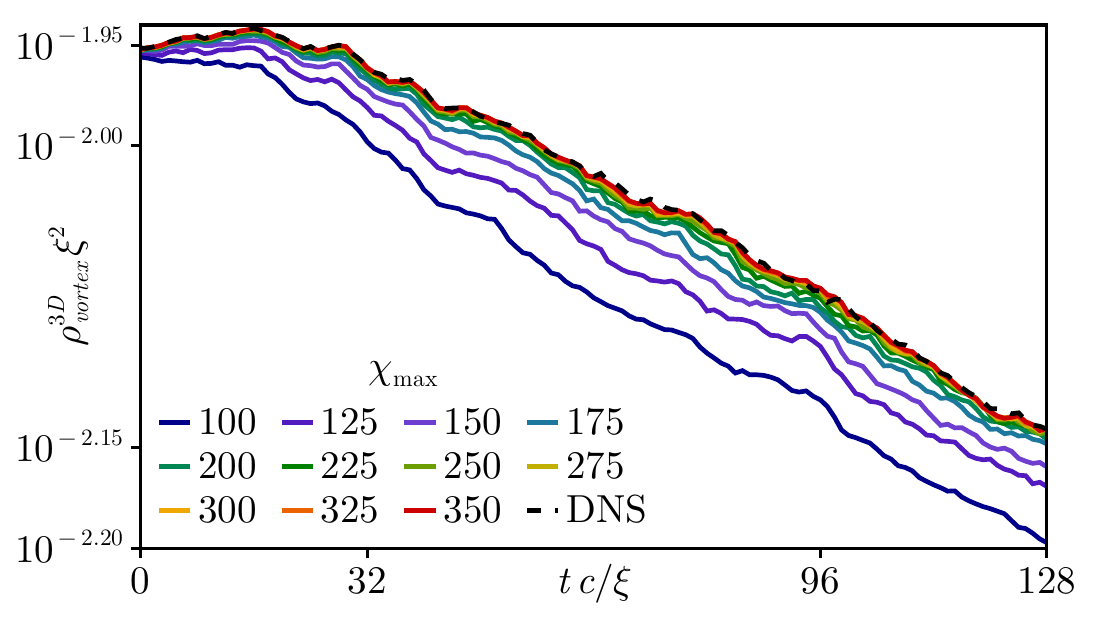}
    \caption{Vortex density $\rho_{\text{Vortex}}^{3D}$ as a function of time $t$ using DNS and MPS with various bond dimensions $\chi_{\text{max}}$ for the simulation of Fig.~\ref{fig:QT3D_Profiles}.}
    \label{fig:QT3D_VortexNumber}
\end{figure}

To quantify the accuracy of the 3D MPS simulations, we focus on the incompressible component and compare the decay rate of the vortex line density $\rho_{\text{Vortex}}^{3D}$ with the DNS results. In Fig.~\ref{fig:QT3D_VortexNumber} we present the density curves for different $\chi_{\text{max}}$ values. Here, we note that $\chi_{\text{max}} \gtrapprox 275$ is sufficient to capture the line density decay, as it suppresses the unphysical damping. We find a slight difference in the $\chi_{\text{max}}$ value obtained in this section compared to the corresponding value in Fig.~\ref{fig:Scaling}(c) of $\chi_{\text{max}} = 285$ (dotted line). Similarly to the 2D case, the corresponding memory ratio ($7\%$) is larger than that of the truncated DNS result in Fig.~\ref{fig:Scaling}(i) ($2\%$, dotted line).

\section{Conclusions}\label{sec:Conclusions}

In this work, we have presented a tensor network-based numerical method to solve the damped GP equation. The use of TDVP for the time evolution of the MPS representation correctly captures the dynamics of the simplest nonlinear excitations, namely dark solitons in 1D, and vortices in 2D and 3D. For these examples, the method requires down to $\sim 1\%$ of the memory that the DNS uses, with the highest memory savings being realized for 3D examples. In particular, we find that the maximum bond dimension for a fixed density of nonlinear excitations of a turbulent state scales as $\chi_{\text{max}} \sim \sqrt{L^d} \sim \sqrt{M}$ for $d$ dimensions and $M$ total number of grid points. This scaling is a specific property of the MPS ansatz. Moreover, it was shown that for large system sizes (of extent $L$), the memory required by MPS is an increasing, monotonous function of the density of the nonlinear excitations. This tendency enables us to achieve a memory reduction of 4 orders of magnitude ($0.03\%$) in the study of vortex line reconnections. Here we capture the principal physical characteristics of vortex line dynamics, such as Kelvin waves, Crow instabilities, and vortex ring cascades.

We also analyze the nontrivial dynamics of multiple nonlinear excitations, namely quantum turbulence. We reproduce the dynamics of the 1D dark soliton gas with infidelities of $I \approx 10^{-5}$ with respect to that of DNS. For 2D, we find that the two-point spatial averaged correlations and the number of vortices are well represented with moderate bond dimensions. For 3D, we confirm that the expected decay of the vortex line density accurately describes the dynamics. This results in MPS simulations using memories of $\sim 27\%$ and $\sim 7\%$ with respect to that which DNS requires for 2D and 3D, respectively. The expected incompressible energy cascade in 2D is recovered for even lower bond dimensions, as the vortex structures are well represented despite the artificial damping generated by an exceeded truncation threshold. The latter implies that, if we are primarily interested in the turbulent statistics of the incompressible component, a more dramatic computational saving is possible with the MPS representation. This fact is significant since the regime of strong (vortex-dominated) turbulence lies beyond the framework of wave turbulence that allows for turbulence closures to be obtained~\cite{Zakharov_2025}. It follows that efficient numerical simulations with favorable scaling properties, such as those based on tensor networks, are essential for modeling these regimes of quantum turbulence. The developments shown here emphasize the importance of tensor networks for solving nonlinear PDEs. These techniques could lead to a paradigm shift in how we tackle specific computationally-intensive simulations with the potential to uncover new multiscale physics.

This work paves the way for running computationally demanding quantum turbulence simulations on accessible hardware, with multiple avenues for future research. The MPS techniques shown here can be expanded to tackle extensions of the GP equation. For instance, generalized Ginzburg–Landau models incorporate thermal effects or noncondensed components~\cite{Griffin_2009, Proukakis_2013} used to study nonequilibrium condensates, including superconductivity~\cite{Aranson_RMP2002} and exciton-polaritons~\cite{Kasprzak_Nat2006, Berloff_PhysicsofQuantumFluids2013}. These MPS methods can also lead to an efficient tensor network representation of the novel supersolid state~\cite{Lahaye_RPP2009, Chomaz_RPP2022, Sanchez-Baena_NatC2023, Casotti_Nat2024, Kirkby_PRR2025}. Moreover, variants of the GP model also provide qualitative insights into other condensed matter systems such as superfluid $\text{He}^4$~\cite{Berloff_PNAS2014, Skrbek_PNAS2021} and ion–vortex complexes~\cite{Berloff_PRB2000, Jin_PRB2010, Villois_PRB2018}. Here, given that systems of superfluid $\text{He}^4$ are highly incompressible and support a much larger range of length scales than what is currently accessible in cold atomic gases, the MPS representation will be particularly relevant. Then, the advantage of the MPS ansatz can be exploited to study more complex sparse vortex configurations, such as vortex bundle reconnections, which are argued to mimic eddies in classical turbulence~\cite{Alamri_PRL2008}. In the aforementioned example applications, being able to post-process the computational results directly within the MPS representation is crucial to retain the efficient compression provided by the MPS ansatz. We have demonstrated how this can be realized by developing a vortex tracking algorithm that is explicitly formulated for the MPS of the wavefunction. This enables the tracking of complex spatio-temporal vortex dynamics, including vortex reconnections. In addition, the developed tracking facilitates the evaluation of essential experimentally measurable diagnostics, such as vortex line densities, from our simulations.

Alternative tensor network architectures could be considered to further improve the scaling properties of the method. In particular, we note that tree tensor networks have already appeared in the literature of the quantics representation~\cite{Ye_JPP2024, Tindall_2024}. Other geometries, such as the projected entangled pair states (PEPS)~\cite{Cirac_RMP2021} and multiscale entanglement renormalization ansatz (MERA)~\cite{Vidal_PRL2008}, are promising tensor networks that remain unexplored for quantum-inspired numerical methods. Additional performance enhancements can be potentially realized with
different tensor-network time evolution methods. These include the use of quantum Fourier transforms for the kinetic term~\cite{Chen_PRXQ2023, Connor_2025}, and the novel tensor cross interpolation methods~\cite{Oseledets_LAA2010, Dolgov_CPC2020, Savostyanov_LAA2014, Ghahremani_CMAME2024, Niedermeier_2025, NunezFernandez_SciPP2025}. The latter allows the nonlinear terms to be implemented with a scaling of $\mathcal{O}(\chi^3)$ instead of the $\mathcal{O}(\chi^4)$ scaling that currently forms the main bottleneck in the implementation of our numerical scheme. Furthermore, a quantum computer can correctly encode the interlength-scale entanglement of the quantics ansatz. Then, the use of shallow circuits to extend the method shown here to quantum hardware~\cite{Jaksch_AIAA2023, Tennie_NatRP2025, Alipanah_2025} could lead to an optimal performance.\\

\emph{Note:} Concurrently with this manuscript, independent works have very recently appeared that also apply tensor network approaches to the GP equation. These works consider different applications: In Ref.~\cite{Bou-Comas_2025}, the focus is on 1D problems including Bose-gas mixtures and spinor BECs; in Ref.~\cite{Chen_2025} the authors illustrate the method for obtaining ground states in 1D and 2D; in Ref.~\cite{Niedermeier_2025} the efficacy of the algorithm is discussed for simulating the dynamics subject to 1D and 2D quasicrystal potentials; and in Ref.~\cite{Connor_2025} 3D simulations are also considered with more emphasis on dipolar BECs.

\acknowledgements

The authors are very grateful to N. Berloff, M. Stoudenmire, R. Pinkston, and H. Alipanah for fruitful discussions. This research was supported in part by the University of Pittsburgh Center for Research Computing and Data, RRID:SCR\_022735, through the resources provided. Specifically, this work utilized the HTC cluster, supported by NIH award number S10OD028483, and the H2P cluster, supported by NSF award number OAC-2117681.

\appendix

\section{Operators on the quantics representation} \label{sec:OperationsQTT}

Here we describe the two primary operations to formulate the damped GP equation (Eq.~\eqref{eqn:dGPE}) in the quantics MPS framework: multiplication between functions and evaluating derivatives. Both are performed using the representation of operations in matrix product structures, namely matrix product operators (MPOs)~\cite{Schollwock_AP2011}. For details on other core MPS operations, such as the sum of MPSs, which can help solve PDEs, we refer the reader to multiple reviews, such as Refs.~\cite{Verstraete_AdvP2008, Schollwock_AP2011, Orus_AP2014, Ran_2020, Catarina_EPJB2023}.

\subsection{Hadamard product} \label{sec:Hadamard}

\begin{figure}[t]
    \centering
    \includegraphics[width=\linewidth]{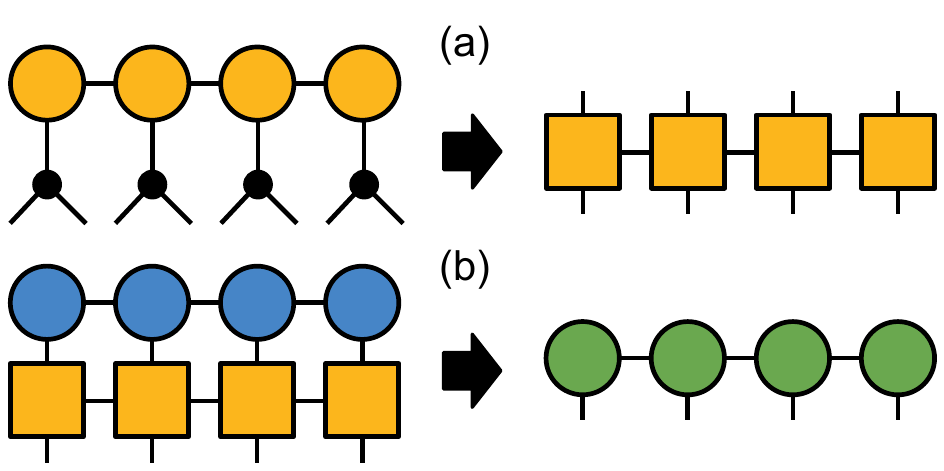}
    \caption{Depiction of the Hadamard product. (a) Promoting an MPS (yellow circles) to an MPO (yellow squares) using 3-rank delta tensors $\delta_{i,j,k}$ (black circles). (b) Applying the resulting MPO to another MPS results in an MPS containing the element-wise product of the two respective states.}
    \label{fig:HadamardProduct}
\end{figure}

Consider an MPS that represents a function using the quantics framework from Sec.~\ref{sec:EncodingMPS}. The element-wise multiplication operation for this format is called the Hadamard product~\cite{Gourianov_2022}, which is performed using 3-rank delta tensors $\delta_{i,j,k}$ that equal 1 for $i=j=k$ and zero otherwise. We apply a delta tensor for each physical index in the MPS, and after contraction, we obtain the corresponding MPO shown in Fig.~\ref{fig:HadamardProduct}(a). This MPO contains the information of the original continuous function as a diagonal operator. Then, applying the MPO to a quantics MPS of a second function results in the element-wise product of the two functions also encoded in the quantics MPS format, a process depicted in Fig.~\ref{fig:HadamardProduct}(b).

Various methods for efficiently performing MPO-MPS products have been proposed~\cite{Stoudenmire_NJP2010, Michailidis_2024, Camano_2025}. Here we consider a variational optimization algorithm, called \emph{fit} in ITensor~\cite{Fishman_SciPPC2022, Fishman_SciPPC2022a}, that resembles the DMRG algorithm~\cite{Schollwock_AP2011}. The \emph{fit} algorithm consists of performing sweeps across the system by optimizing each tensor individually while keeping the remaining tensors fixed. The computational complexity of the method scales as $\mathcal{O}(N \eta \chi^3)$ where $\eta$ is the bond dimension of the MPO, $\chi$ is the bond dimension of the MPS, and $N$ is the number of physical indices. To calculate $|\psi|^2$ using the Hadamard product, we have $\eta = \chi$, hence the cost scales as $\mathcal{O}(N \chi^4)$. As an initial guess for the variational optimization, we use the MPS of $|\psi|^2$ from the previous timestep. Thus, only one sweep is needed to obtain a good approximation. The resulting bond dimension can grow up to $\chi^2$; hence, the algorithm performs truncation during sweeping to prevent uncontrolled growth of the bond dimension.

\subsection{Derivatives} \label{sec:Derivatives}

The derivatives are constructed using an eighth-order finite difference approximation with periodic boundary conditions~\cite{Fornberg_MC1988}. To build the latter, we first write the shift operators (translation operators) over one grid spacing $h$, to the left $\hat{L}f(x) = f(x+h)$ and right $\hat{R}f(x) = f(x-h)$, as MPOs. For a quantics MPS of a one-dimensional function, the shift operators have an exact MPO form with a bond dimension $\chi = 2$~\cite{Kazeev_SIAMJMAA2012} given by
\begin{align}
    \hat{R} = 
    \begin{bmatrix}
        1 & 1
    \end{bmatrix}
    \prod_{n=1}^N\begin{bmatrix}
        \hat{I} & \hat{a}^\dagger_n \\
        \hat{0} & \hat{a}_n
    \end{bmatrix}
    \begin{bmatrix}
        0 \\ 1
    \end{bmatrix}, \label{eqn:rigth_shift_MPO} \\
    \hat{L} = 
    \begin{bmatrix}
        1 & 1
    \end{bmatrix}
    \prod_{n=1}^N\begin{bmatrix}
        \hat{I}_n & \hat{a}_n \\
        \hat{0}_n & \hat{a}^\dagger_n
    \end{bmatrix}
    \begin{bmatrix}
        0 \\ 1
    \end{bmatrix}. \label{eqn:left_shift_MPO}
\end{align}
Here $\hat{I}_n$ ($\hat{0}_n$) is the $2\times2$ identity (zero) operator, and $\hat{a}_n$ ($\hat{a}^\dagger_n$) is the annihilation (creation) operator, acting on the $n$-th qubit according to Eq.~\eqref{eqn:QTT}.

We extend the shift operator definitions to higher dimensions by first expanding Eqs.~\eqref{eqn:rigth_shift_MPO} and \eqref{eqn:left_shift_MPO} as
\begin{align}
    \hat{R}^{(d)} &= \sum_{l=1}^{N} \left( \prod_{m=N-l+2}^{N} \hat{a}^{(d)}_{m} \right) \hat{a}^{(d)\dagger}_{N-l+1} + \prod_{m=1}^{N} \hat{a}^{(d)}_{m}, \label{eqn:right_shift} \\
    \hat{L}^{(d)} &= \sum_{l=1}^{N} \left( \prod_{m=N-l+2}^{N} \hat{a}^{(d)\dagger}_{m} \right) \hat{a}^{(d)}_{N-l+1} + \prod_{m=1}^{N} \hat{a}^{(d)\dagger}_{m}. \label{eqn:left_shift}
\end{align}
In this case, the $\hat{a}^{(d)}_n$, $\hat{a}^{(d)\dagger}_n$ operators correspond to the $n$-th qubit of the $d$-axis ($x$, $y$, or $z$) according to the different orderings shown in Fig.~\ref{fig:MPSOrder}. Then, several algorithms can be used to create the MPO based on Eqs.~\eqref{eqn:right_shift} and \eqref{eqn:left_shift} that achieve the minimal bond dimension ($\chi = 2$ for 1D, for example )~\cite{Ren_JCP2020, Chan_JCP2016, Hubig_PRB2017, Keller_JCP2015, Ehlers_PRB2017, Corbett_2025}. Using the explicit expressions from
Eqs.~\eqref{eqn:right_shift} and \eqref{eqn:left_shift} enable the testing of multiple MPS orders without requiring manual MPO construction.

Using the shift operators, we can build finite difference formulas of arbitrary order by applying powers of operators $\hat{R}$ and $\hat{L}$. For instance, $f(x-2h)$ corresponds to the MPO-MPO product $\hat{R} \cdot \hat{R}$. Using a tolerance $\varepsilon = 10^{-16}$ (that is close to machine precision), we obtain for the eighth-order finite difference Laplacian the maximum bond dimensions shown in Table~\ref{tab:LaplacianBonds}, most of which are of $\mathcal{O}(1)$ for arbitrary system size. 
\begin{table}[t]
    \centering
    \begin{tabular}{llll}
        \hline
         & 1D & 2D & 3D \\ \hline
        Sequential & 3 & 4 & 4 \\
        Staircase &  & 4 & 4 \\
        Interleaved &  & 6 & 8 \\ \hline
    \end{tabular}
    \caption{Bond dimensions $\chi$ of the Laplacian $\nabla^2$ MPO representation in one to three dimensions using the sequential, staircase, and interleaved orders depicted in Fig.~\ref{fig:MPSOrder}. We use $\varepsilon = 10^{-16}$ and saturation with the MPS size was checked for a large number of qubits ($\leq 100$ qubits per axis).}
    \label{tab:LaplacianBonds}
\end{table}
We note that recent work has shown that eighth-order formulas have similar errors in calculating derivatives compared to a spectral method using the quantum Fourier transform~\cite{Connor_2025}.

\section{Initial conditions}\label{sec:InitialConditions}

Below, we provide the details for constructing the initial conditions that are used throughout the article.

\subsection{1D Dark soliton}\label{sec:DarkSoliton_Initial}

Using the exact result from Ref.~\cite{Sato_NJP2016} we build the dark soliton ansatz $\psi(x) = \sqrt{\rho(x)}e^{i\phi(x)}$ corresponding to
\begin{align} 
    \rho(x) &= \rho_1 + (\rho_0 - \rho_1) cn\left(u\left(\frac{x}{L};k\right), k\right)^2, \label{eqn:DarkSoliton1} \\
    \phi(x) &= vx - \left( \frac{Lv}{2} + \pi \right) \frac{\Pi\left[ 1 - \frac{\rho_1}{\rho_0}, am\left( u \left(\frac{x}{L};k \right), k \right),  k \right]}{\Pi\left[ 1 - \frac{\rho_1}{\rho_0}, k \right]}. \label{eqn:DarkSoliton2}
\end{align}
Here $cn(x,k)$ and $am(x,k)$ are the elliptic cosine and Jacobi amplitude, $\Pi(n,k)$ and $\Pi(n; \phi | k)$ are the complete and incomplete elliptic integrals of the third kind, and $u(t;k) = K(k) \left( 2t-1 \right)$ where $K(k)$ is the complete elliptic integral of the first kind. The solution is parametrized by the constants $\rho_0$, $\rho_1$, $k$, and $v$. The periodic boundary conditions define the following restrictions~\cite{Sato_NJP2016}:

\begin{figure}[b!]
    \centering
    \includegraphics[width=\linewidth]{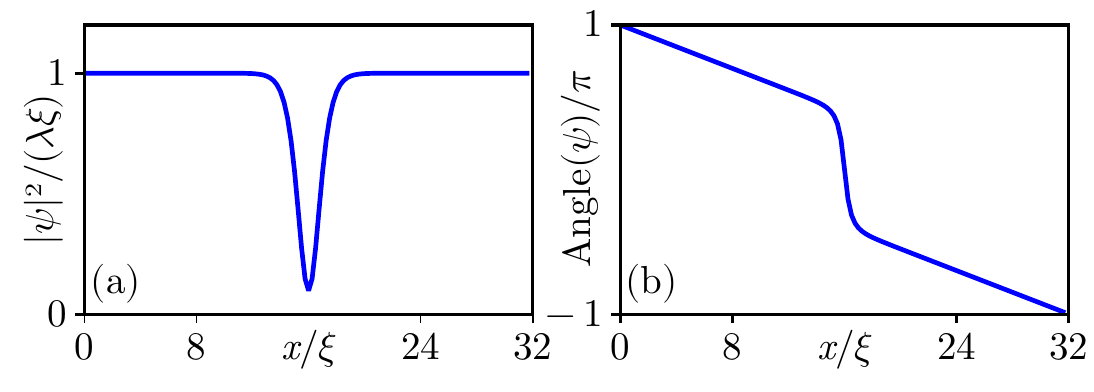}
    \caption{Density profile $|\psi|^2$ (a) and phase $\text{Angle}(\psi)$ (b) of the dark soliton initial condition with periodic boundary conditions using $L = 32 \, \xi$ and $k = 1 - 5 \times 10^{-13}$.}
    \label{fig:DarkSolitonInitialCondition}
\end{figure}

\begin{align}
    \rho_1 &=\rho_0 + \left(\frac{2kK(k)}{L} \right)^2, \label{eqn:DarkSoliton3} \\
    v &= \pm \frac{2}{L}\sqrt{\frac{\rho_1}{\rho_0} \left( 1 + \frac{k^2}{\frac{\rho_1}{\rho_0} - 1} \right)} \Pi\left(1 - \frac{\rho_0}{\rho_1}, k \right) + \frac{2\pi n}{L}, \label{eqn:DarkSoliton4}
\end{align}
where $n$ is an integer. We choose the first term to have a positive sign and $n = -1$ such that we add a phase gradient $2\pi/L$ to introduce a $2\pi$ phase jump across the system. For the example shown in Fig.~\ref{fig:DarkSoliton}, we fix $\rho_1 = 1$ and $k = 1 - 5 \times 10^{-13}$, which defines the rest of the parameters and results in a velocity of $v \approx 0.19 \, c$. The initial condition is depicted in Fig.~\ref{fig:DarkSolitonInitialCondition}. For comparison with the numerical simulations in Sec.~\ref{sec:DarkSoliton}, we use the exact evolution of the dark solution given by $\psi(x,t) = \sqrt{\rho(x-vt)}e^{i\phi(x-vt)}$ up to a time-dependent phase factor.

\subsection{2D Vortex dipole}\label{sec:VortexDipole_Initial}

\begin{figure}[t]
    \centering
    \includegraphics[width=\linewidth]{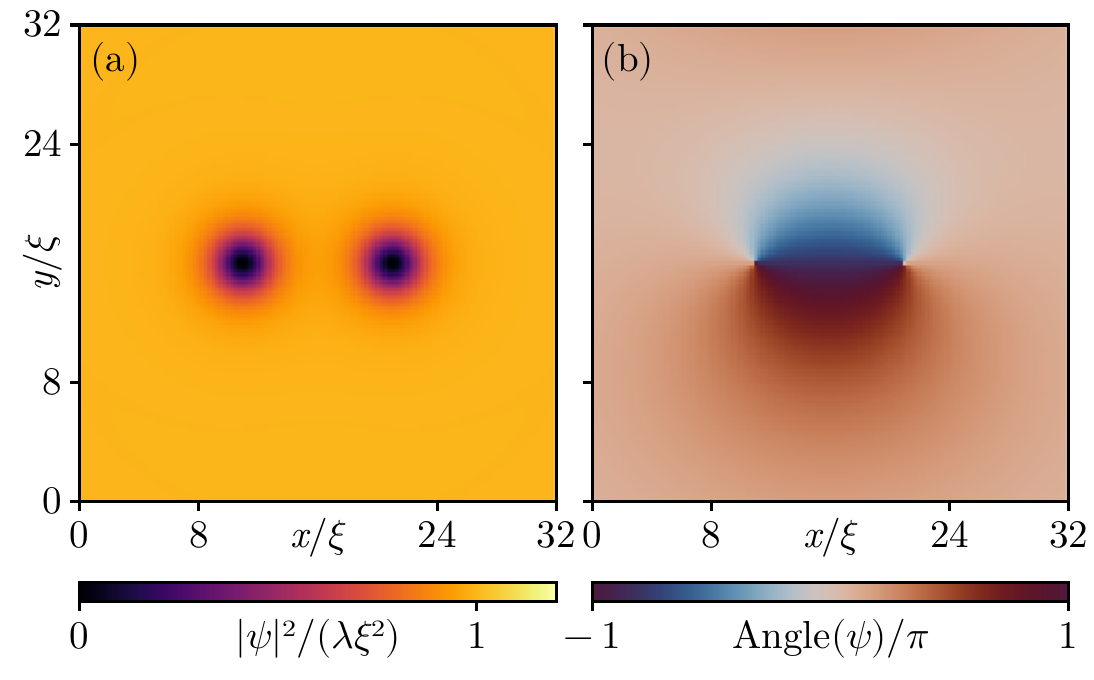}
    \caption{Density profile $|\psi|^2$ (a) and phase $\text{Angle}(\psi)$ (b) of the vortex dipole initial condition with periodic boundary conditions using $L = 32 \, \xi$ and $d = 10 \, \xi$.}
    \label{fig:VortexDipole}
\end{figure}

Here, the wave function consists of the composition of an ansatz for each vortex, both with opposite charges, given by
\begin{align}
    \psi(\vb{r}) &= 
    \psi_1(\vb{r}) \cdot \psi^*_2(\vb{r}) \cdot e^{i(\theta_1(x,L) - \theta_2(x,L)) \frac{2y}{L}}, \label{eqn:VortexDipole1} \\
    \psi_j(\vb{r}) &= \sqrt{\frac{|\vb{r}-\vb{r}_j|^2(a_1 + a_2 |\vb{r}-\vb{r}_j|^2)}{1+b_1 |\vb{r}-\vb{r}_j|^2 + b_2 |\vb{r}-\vb{r}_j|^4}} \, e^{i\theta_j(\vb{r})},  \\
    a_1 &=  11/32, \label{eqn:VortexDipole2}\\
    a_2 &= a_1(b_1-1/4), \label{eqn:VortexDipole3}\\
    b_1 &= \frac{5 - 32 a_1}{48 - 192 a_1},  \label{eqn:VortexDipole4}\\
    b_2 &= a_2, \label{eqn:VortexDipole5}
\end{align}
where $\vb{r} = (x,y)$ is the position vector and $\vb{r}_j$ corresponds to the initial position of the $j$-th vortex core. The vortex dipole is characterized by a dipole length $d = |\vb{r}_2-\vb{r}_1|$, and $\theta_j(\vb{r}) = \text{Angle}(\vb{r}-\vb{r}_j, \vb{r}_2-\vb{r}_1)$ is the angle of $\vb{r}-\vb{r}_j$ measured from the direction $\vb{r}_2-\vb{r}_1$. Here, $\psi_j$ is the single vortex core ansatz obtained from Padé approximations~\cite{Berloff_JPA2004}. Since we are using periodic boundary conditions, a phase gradient is added that matches the upper and lower boundaries of the wave function. In Fig.~\ref{fig:VortexDipole} we show an example for $L = 32 \, \xi$ where we locate the positive charge at $\vb{r}_1 = ((L-d)/2, L/2)$ and the negative one at $\vb{r}_2 = ((L+d)/2, L/2)$, with a dipole lenght $d = 10\, \xi$. The dipole moves upwards as shown in Sec.~\ref{sec:VortexDipole}.

\subsection{3D Vortex ring}\label{sec:VortexRing_Initial}

\begin{figure}[t]
    \centering
    \includegraphics[width=0.5\linewidth, trim={1.5cm 0.5cm 3.5cm 4.0cm}, clip]{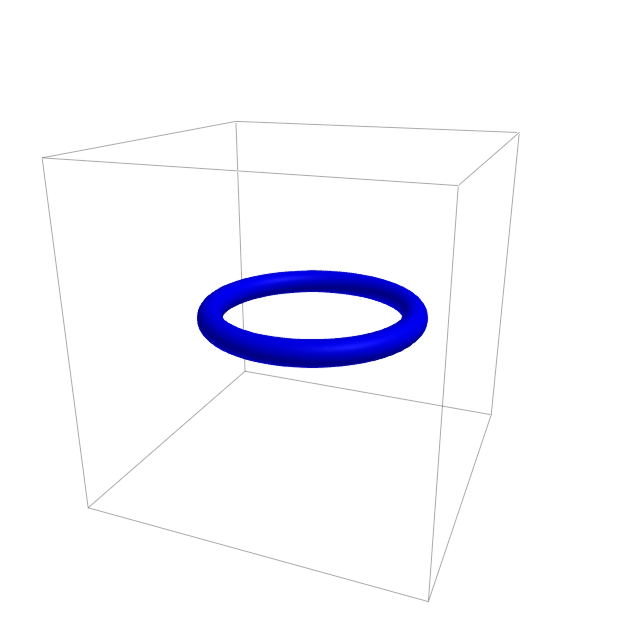}
    \caption{Initial condition for the vortex ring with periodic boundary conditions using $L = 32 \, \xi$ and $R = 5 \, \xi$. We show the contour at a density $|\psi|^2 = 0.1 \, \lambda \xi^3$ in the box $[L/4, 3L/4]^3$.}
    \label{fig:VortexRing}
\end{figure}

The vortex ring is generated by taking the ansatz from Appendix~\ref{sec:VortexDipole_Initial} and rotating it with respect to the axis along which the dipole moves. Then, for a given radius $R$ we define $\vb{r}_1 = (-R,L/2)$, $\vb{r}_2 = (R,L/2)$ and $\vb{r}=(\sqrt{(x-L/2)^2+(y-L/2)^2}, z)$, and substitute in Eqs.~\eqref{eqn:VortexDipole1}-\eqref{eqn:VortexDipole5} resulting in a dipole length $d = 2R$ and a direction of movement along the positive $z$-axis. The correction for periodic boundary conditions in the 2D dipole ansatz also applies to this 3D case. We show an example for $L = 32 \, \xi$ and $R = 5\,  \xi$ in Fig.~\ref{fig:VortexRing}.

\subsection{Vortex line reconnection}\label{sec:VortexLineReconnection_Initial}

\begin{figure}[b
]
    \centering
    \includegraphics[width=\linewidth]{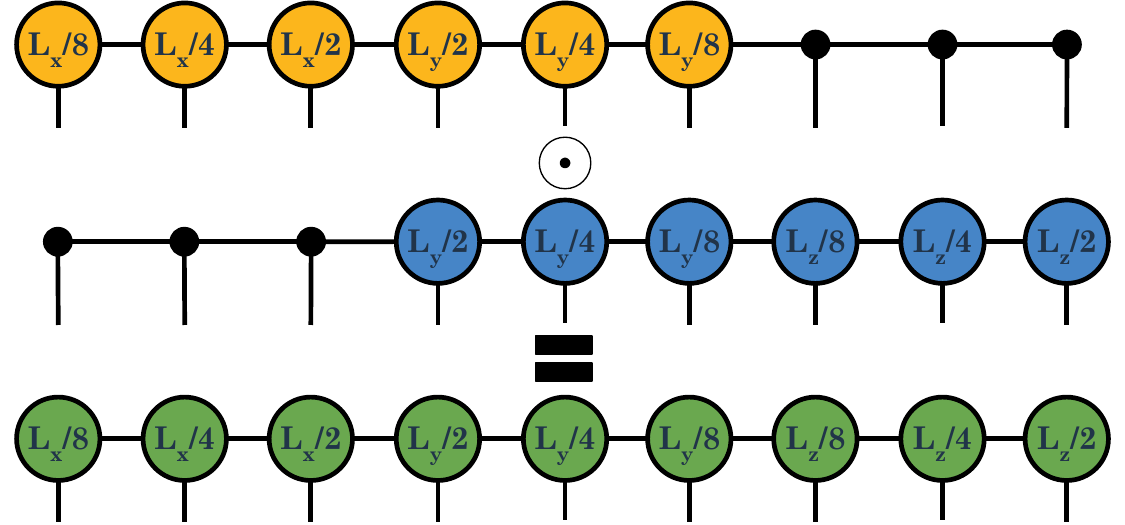}
    \caption{3D extension of 2D quantics MPS. The yellow and blue MPS correspond to quantics representations of 2D functions (e.g. vortex dipoles), in the $xy$ and $yz$ planes, respectively. Delta tensors $\delta_{ijk}$ (bluk) and $\delta_{ij}$ (boundaries) are drawn as black circles. The $\odot$ indicates the Hadamard product of the two MPS, resulting in the green 3D quantics MPS. The corresponding scales of each qubit are indicated with the convention from Fig.~\ref{fig:MPSOrder}.}
    \label{fig:MPS2DTo3D}
\end{figure}

Each dipole is initialized in a 2D quantics MPS by using the initial condition from Appendix~\ref{sec:VortexDipole_Initial} and applying iterative SVDs. Also, an imaginary time evolution is applied to suppress initial sound waves. The 2D quantics MPS is then extended to a 3D quantics MPS by adding delta tensors for the missing qubits that represent the third coordinate direction. This process is depicted in Fig.~\ref{fig:MPS2DTo3D}, where initial 2D quantics MPS in the $xy$ (yellow) and $yz$ (blue) planes have attached delta tensors (black circles) that account for the missing spatial dimension. Then, a Hadamard product of the two MPS results in the 3D quantics MPS (green) representing the four vortex lines. The initial 2D quantics MPS have different staircase orderings, one coupled through the large length scales ($xy$ plane) and the other through the small scales ($yz$ plane). This results in a 3D quantics MPS with a staircase $xyz$ order as shown in Fig.~\ref{fig:MPSOrder}(d). The latter choice minimizes the correlations at the edges of the final-time MPS. Instead, the initial orderings focus the most significant correlation contribution at the center of the MPS in the overlapping region, where more singular values are available according to the iterative SVD construction of an MPS (see Sec.~\ref{sec:EncodingMPS}).

\subsection{Random phase interpolation}\label{sec:RandomPhaseInterpolation}

Motivated by Kobayashi, \textit{et al.}~\cite{Kobayashi_PRL2005}, we first create a coarse grid with $N_{\text{RPI}} < N$ qubits per axis, choosing randomly-generated values of the wave function on each point. For the DNS, we sample uniformly from the interval $[-1, 1]$ for both the real and imaginary components. The sampling region needs to cross zero, allowing phase jumps to appear after interpolation. In the quantics MPS case, we randomly choose the entries of the MPS tensors with given maximum bond dimension $\chi_{\text{max}}^{\text{RPI}}$. In contrast to the DNS case, the MPS initial condition samples the wave function from nonuniform distributions. These are generated from linear combinations of the random components of the MPS after contracting the tensor network, whose real and imaginary parts are distributed uniformly within the interval $[-1,1]$. Even then, these nontrivial random distributions are still symmetric with respect to zero; hence, phase jumps emerge. We typically set a small value of $\chi_{\text{max}}^{\text{RPI}} \leq 32$, as its variation only results in a different vortex distribution but does not significantly affect the accuracy of the results.

After generating the initial state, successive linear interpolations are performed until a resolution with $N$ qubits per axis is achieved. The linear interpolation in the quantics MPS format is performed as follows: First, given an MPS with $N$ qubits per axis, for each one, we add a qubit next to the current smallest length scale with the state $\ket{0}+\ket{1}$, equivalent to an identity in the quantics MPS formalism. If the new qubit is at the edge of the MPS, we add to it a trivial bond index with dimension $1$. On the other hand, if it has to be added in the middle of the chain, we split the corresponding bond index $c$ into two equivalent copies $l$, $r$. Afterwards, we create a semi-diagonal tensor connected to both bonds, defined with an identity state as $\ket{0}+\ket{1}$ for $l=r$ and zero otherwise. Then, a linear projection operator $\hat{P}$ is applied to the respective qubits for each axis. This operator is defined as
\begin{align}
    \hat{P} = \hat{I} - \frac{1}{2}\sum_{n = 1}^N \hat{P}^n_{\ket{0}+\ket{1}} \cdot \prod_{m = n+1}^{N} \hat{a}^\dagger_m  \cdot \hat{P}^{N+1}_{\ket{0}},
\end{align}
where $\hat{I}$ is the identity operator, $\hat{P}^n_{\ket{u}} = \dyad{u}$ is the projection operator of the state $\ket{u}$, $\hat{a}_n^\dagger$ is the creation operator and the index $n$ indicates the $n$-th qubit of the corresponding axis ordered from the largest to the smallest length scale. For 1D, this operator has an exact low-rank MPO with bond dimension $\chi_{\text{max}} = 2$~\cite{Gourianov_2022}. This expression is straightforward to apply to higher-dimensional quantics MPS by keeping track of the corresponding length scale order, analogous to Eqs.~\eqref{eqn:right_shift} and \eqref{eqn:left_shift}. To reach a desired number of qubits, the previous procedure is iterated such that, at each time, the number of qubits per axis is increased by one.

Once the state is prepared as described above, it is integrated forward in imaginary time to ensure the respective excitations are imprinted while damping sound waves from the system. For large enough times, the imaginary time evolution drives the system to its ground state, that is, a uniform density. To avoid this, we remove the projection of the initial random state onto the ground state with a Gram-Schmidt process. The latter can also be performed directly with MPS operations, as it only requires linear transformations. In 1D, the dark solitons are not topologically protected; hence, both the dark solitons and the compressible sound waves decay upon integrating the system in imaginary time for extended durations. Therefore, in the 1D case, we adopt the DNS and fix the initial phase of the system during the imaginary time evolution. This procedure generates dark solitons around phase jumps along the chain. We use cubic splines with periodic boundary conditions for interpolation instead of a linear ansatz, allowing us to recover a smooth initial wave function. For 2D and 3D, the vortices are topologically protected. This fact implies that under imaginary time evolution, the vortices only disappear if a pair of oppositely signed vortices (or anti-parallel vortices in 3D) annihilate each other's phase singularities. For this case, it is sufficient to use linear interpolation during the initialization stage, since the imaginary time evolution smooths the wave function.

\begin{figure}[t]
    \centering
    \includegraphics[width=\linewidth]{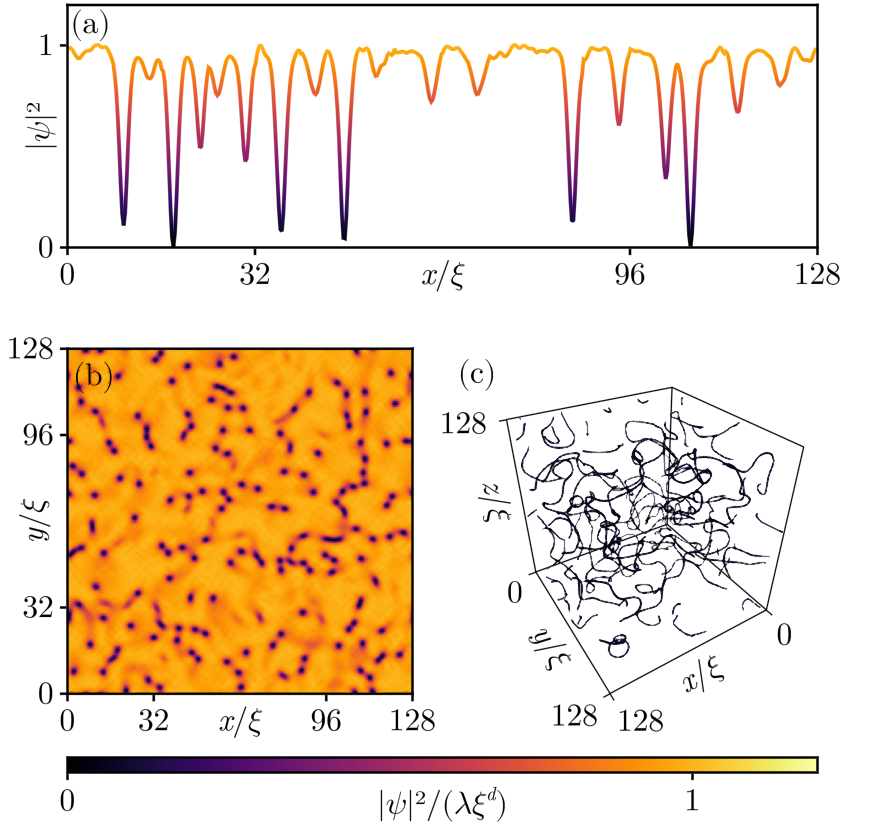}
    \caption{
        Density profiles $|\psi|^2$ of integrable (1D) and nonintegrable (2D, 3D) quantum turbulence states generated using the modified random phase interpolation. (a) 1D Dark soliton. (b) 2D point-vortex distribution. (c) 3D vortex tangle, where we show the density at $|\psi|^2 = 0.1 \, \lambda \xi^3$. For each simulation in $d$ number of spatial dimensions, the domain size is $L = 128\,\xi$ along each coordinate direction, and we use $9$ qubits per axis.
    }
    \label{fig:TurbulenceGeneration}
\end{figure}

In Fig.~\ref{fig:TurbulenceGeneration} we show an example in each dimension for the generation of a turbulent state, all with domains of length $L = 128\, \xi$ and $9$ qubits per axis, corresponding to a resolution of $\xi/4$. In panel (a) we show an initial soliton gas with density $\rho_{\text{Soliton}} = 8.3 \times 10^{-2} \, \xi^{-1}$ which an MPS well represents with $\chi_{\text{max}} = 8$ (up to an infidelity of $I \approx 10^{-5}$). We also build quantum turbulence states of vortex densities $\rho_{\text{Vortex}}^{2D} = 9.0 \times 10^{-3} \, \xi^{-2}$ (panel (b)) and $\rho_{\text{Vortex}}^{3D} = 3.6 \times 10^{-3} \, \xi^{-2}$ (panel (c)) using $\chi_{\text{max}} = 90$ and $320$, respectively. The two cases correspond to a memory usage of $17\%$ in 2D and $1\%$ in 3D with respect to the DNS. These values are comparable to the ones obtained for the single nonlinear excitations considered in Sections~\ref{sec:VortexDipole} and \ref{sec:VortexRing}.

\section{Length scale orders in quantics MPS} \label{sec:OrderingsMPS}

In this section, we present a detailed analysis of the correlations across the MPS to characterize which length scale ordering works best for the vortex excitation examples considered for 2D (vortex dipole, Sec.~\ref{sec:VortexDipole}) and 3D (vortex ring, Sec.~\ref{sec:VortexRing}). We use this as a basis to select the optimum ordering in the more complex quantum turbulence simulations in Sec.~\ref{sec:QuantumTurbulence}.

\subsection{2D Vortex dipole} \label{sec:Orderings2D}

\begin{figure}[t]
    \centering
    \includegraphics[width=\linewidth]{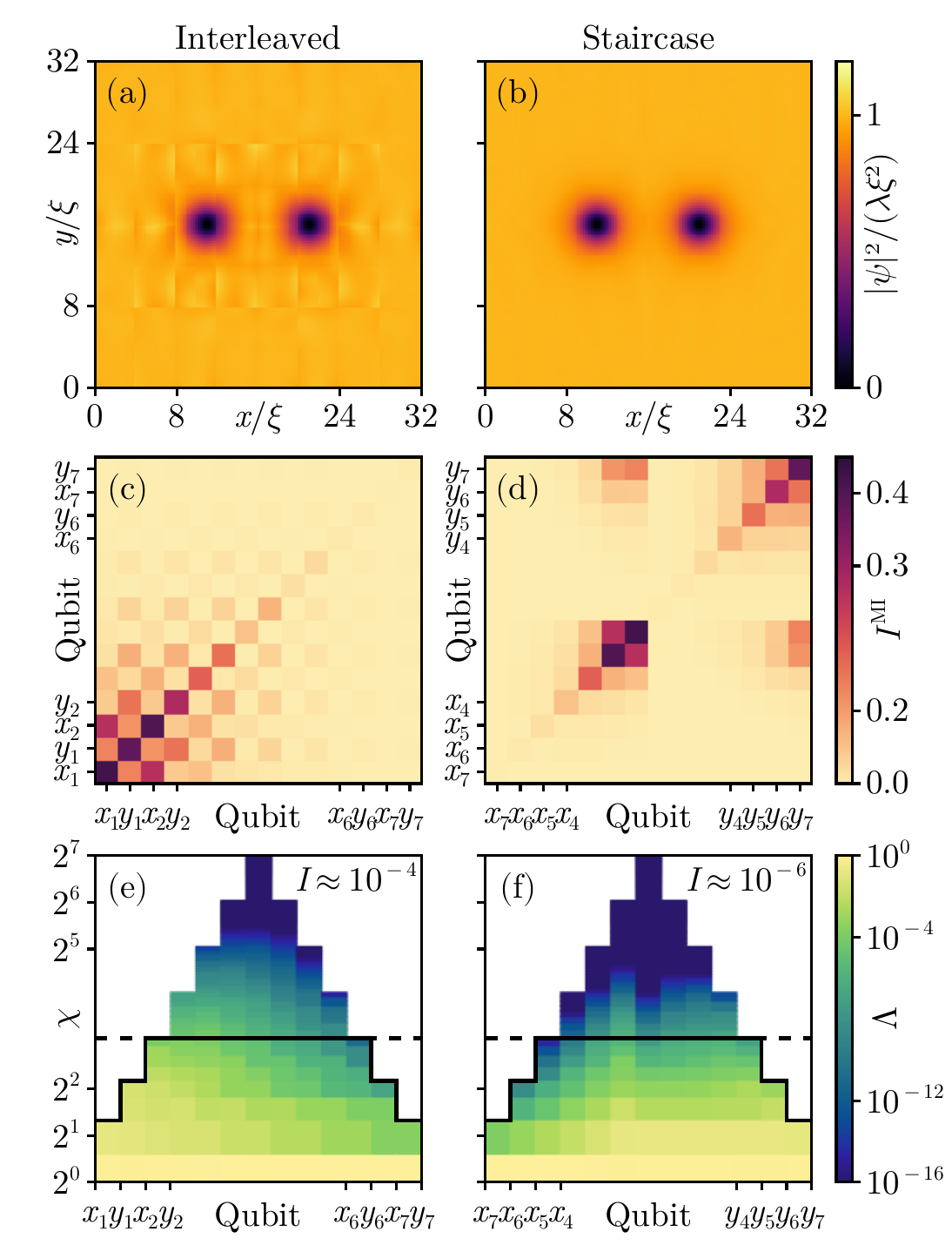}
    \caption{Comparison of vortex dipole quantics MPS representation between interleaved (left) and staircase (right) length scale orderings. (a,b) Density profile of the vortex dipole with a domain size $L = 32\,\xi$ and a dipole length $d = 10\,\xi$ encoded in a system of $14$ qubits ($7$ per axis) truncated with $\chi_{\text{max}} = 8$. (c,d) Two-site mutual information $I^{\text{MI}}$ between length scales of the system with $x_n$, $y_n$ indicating the qubit of the $L \cdot 2^{-n}$ length scale for the corresponding axis. (e,f) Schmidt spectrum $\Lambda$, the black line indicates the bond dimensions $\chi$ across the MPS, and the dashed line indicates $\chi_{\text{max}}$. The corresponding infidelity $I$ with respect to DNS is written in the top right corner.}
    \label{fig:VortexDipoleSpectrum}
\end{figure}

We compare in Fig.~\ref{fig:VortexDipoleSpectrum} the density profiles of a vortex dipole after truncating their MPS representation with the interleaved (a) and staircase (b) orderings, the two possibilities considered in this work for 2D. We use $\chi_{\text{max}} = 8$. Here we observe that, for the same level of truncation, the interleaved ordering exhibits distortions at a scale of $4\,\xi$. On the other hand, the staircase ordering has a smoother density profile around the location of the vortices.

To further illustrate the difference between the interleaved and staircase encodings, we quantify correlations between pairs of length scales (i.e., between qubits) using the two-site mutual information $I^{\text{MI}}$. This measure of correlation is defined by
\begin{align} \label{eqn:mutualinformation}
    I^{\text{MI}}_{l,m} = S_l + S_m - S_{l,m}.
\end{align}
Here, for a given state $\ket{\psi}$ represented by an MPS, $S_l$ is the von Neumman entropy of the reduced density matrix $\rho_l = \text{Tr}_{m \neq l} \dyad{\psi}$ where we trace out all the qubits except for the $l$-th site. Similarly, $S_{l,m}$ is the corresponding entropy for the reduced density matrix of sites $l,m$. We show the two-site mutual information in Fig.~\ref{fig:VortexDipoleSpectrum}(c,d) for each ordering and note that the large scales are highly correlated with each other. In contrast, the small scales are essentially uncorrelated with different scales.
The reason is that the structure of the wave function is mainly characterized by the vortex distribution, which is encoded in the large length scales. On the contrary, the small scales contain the vortex core information, which is the same for all vortices. Hence, these scales do not show any correlation with other structures in the wave function. 

The Schmidt spectrum quantifies the correlations across each bond in the MPS. This spectrum corresponds to a collection of the squared singular values $\Lambda_{j,l} = S_{j,l}^2$ where $S_{j,l}$ is the $j$-th singular value in decreasing order at the $l$-th bond after normalizing as $\sum_j \Lambda_{j,l} = 1$. We plot the corresponding Schmidt spectrum in Fig.~\ref{fig:VortexDipoleSpectrum}(e), which indicates that for the interleaved ordering, the most significant correlations accumulate at the edge of the MPS. These correlations spread through the whole spectrum, resulting in an infidelity of $I \approx 10^{-4}$. In contrast, for the staircase ordering in panel (f), correlations are concentrated at the center of the MPS. Consequently, the rest of the spectrum decays isotropically as the length scale decreases, which produces an infidelity that is two orders of magnitude lower ($I \approx 10^{-6}$). This result is consistent with Ref.~\cite{Connor_2025}, which argues that correlations of the 2D quantum Fourier transform are higher in the interleaved than the sequential ordering (see Fig.~\ref{fig:MPSOrder}(a)).

\subsection{3D Vortex ring} \label{sec:Orderings3D}

Motivated by the results in Fig.~\ref{fig:VortexDipoleSpectrum}, we focus our analysis on the three-dimensional staircase ordering shown in Fig.~\ref{fig:MPSOrder}(d). This MPS ordering has six different possibilities indicated by the various possible permutations of the axes ($x$, $y$, or $z$) across the MPS ($xyz$, $yzx$, $zxy$, $xzy$, $yxz$, $zyx$) while keeping the intra-axis order of the length scales according to Fig.~\ref{fig:MPSOrder}(d). Here, we analyze the representation of the vortex ring with orderings $xyz$, $zxy$, and $yzx$, since the other three are each equivalent to one of the former, given the $xy$ mirror symmetry of the ansatz.

\begin{figure}[t]
    \centering
    \includegraphics[width=\linewidth]{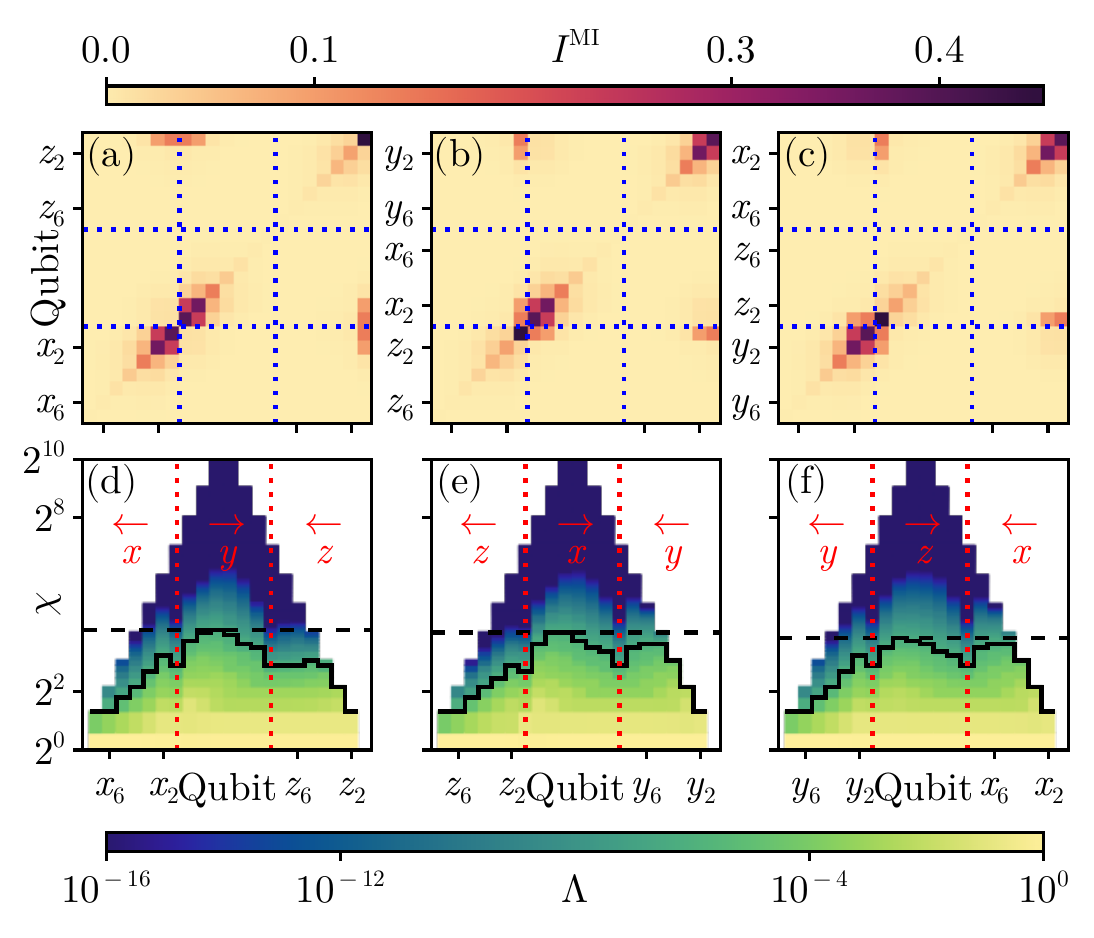}
    \caption{Two-site mutual information $I^{\text{MI}}$ (a-c) and Schmidt spectrum $\Lambda$ (d-f) for an MPS with staircase ordering and different permutations for the ordering of the axes (see Fig.~\ref{fig:MPSOrder}(d)). The dotted blue (red) lines separate the sections for each axis, also indicated in red for (d-f). The arrow signals the order of the qubits from largest to smallest length scales, with $x_n$, $y_n$, $z_n$ indicating the qubit of the $L \cdot 2^{-n}$ scale for the corresponding axis. In (d-f), the black contour indicates the bond dimensions $\chi$ after truncating the state with $\varepsilon = 10^{-6}$; the horizontal dashed line marks the maximum bond dimension in each case. The state represents a vortex ring of radius $R = 5\,\xi$ in a box of length $L = 32\,\xi$ discretized using $3\times 7$ qubits.}
    \label{fig:VortexRingSpectrum}
\end{figure}

With a vortex ring of radius $R = 5 \, \xi$ and $L = 32 \, \xi$, we show the two-site mutual information $I^{\text{MI}}$ in Fig.~\ref{fig:VortexRingSpectrum}(a-c) using $7$ qubits per axis ($128^3$ grid points). 
For all cases, we observe that the highest correlations arise from the largest length scales, analogous to the vortex dipole in Fig.~\ref{fig:VortexDipoleSpectrum}. For the $xyz$ ordering in panel (a), we note that the $xy$ correlations contribute more than the corresponding $xz$ and $yz$ correlations, indicating the complexity of the vortex ring wave function through $xy$ planes. We compare with the corresponding Schmidt spectrum in Fig.~\ref{fig:VortexRingSpectrum}(d) and the bond dimensions (black lines) after truncation with $\varepsilon = 10^{-6}$, for which all orders reach an infidelity $I \approx 10^{-6}$. The large $xy$ correlations generate higher weights in the middle axis than at the edges of the MPS, with a maximum bond dimension (dashed line) of $\chi_{\text{max}} = 14$. The $zxy$ and $yzx$ orderings have similar mutual information distributions (Fig.~\ref{fig:VortexRingSpectrum}(b,c)). However, in these cases, there is a higher contribution from the edge qubits, which balances the corresponding bond dimensions (Fig.~\ref{fig:VortexRingSpectrum}(e,f)). This results in a reduction of the maximum bond dimension that is equal to $\chi_{\text{max}} = 13$ ($zxy$) and $11$ ($yzx$). Since the difference between $\chi_{\text{max}}$ values is small, the analysis of the vortex ring in Sec.~\ref{sec:VortexRing} does not vary appreciably by changing the ordering. Hence, we have opted to work with the $xyz$ staircase ordering to study the vortex ring dynamics presented in the main text.

\section{Vortex line tracking}\label{sec:VortexTracking}

Here, we detail the adaptation of the vortex-tracking method from Ref.~\cite{Villois_JPA2016} using the MPS ansatz. This algorithm enables the extraction of vortex line data from the quantics MPS without resorting to contracting the MPS and explicitly working with extensive grid sizes (e.g., $2048^3$ in Sec.~\ref{sec:VortexLineReconnection}). First, in Appendix~\ref{sec:VortexTracking1} we detail the steps of the algorithm, noting the changes needed and the advantages of using the MPS adaptation during the vortex tracking algorithm. In particular, an important step will be the Fourier interpolation, whose process using MPS we detail after in Appendix~\ref{sec:VortexTracking2}. Finally, in Appendix~\ref{sec:VortexTracking3} we discuss the computational cost of the method. 

\subsection{General algorithm}\label{sec:VortexTracking1}

To obtain a single vortex line, we follow the next steps:
\begin{enumerate}
    \item Choose an initial point in the grid $\vb{r}_0$. The advantage of the MPS ansatz is that it can be interpreted as a probability distribution of its free index entries, for which efficient sampling algorithms have been developed~\cite{Stoudenmire_NJP2010, Ferris_PRB2012}. In particular, points close to a vortex line are characterized by a nonzero pseudovorticity field $\boldsymbol{\omega} = \frac{\hbar}{2i} \nabla\psi^* \times \nabla \psi$. Hence, we build the quantics MPS representing $|\boldsymbol{\omega}|^2$ by first obtaining the derivatives $\nabla \psi$ (one MPS for each component) using Fourier-based methods as explained at the end of this section. Then, we evaluate $\boldsymbol{\omega}$ and its magnitude using the corresponding additions, subtractions, and multiplications with the Hadamard product. With $|\boldsymbol{\omega}|^2$ represented as an MPS, we can efficiently sample optimal starting points close to the vortex lines. We discard values with $|\psi| / \sqrt{\lambda \xi^3} \geq \epsilon$ for a small value of $\epsilon$, which we set to $\epsilon = 0.1$, to improve convergence of the vortex tracking~\cite{Villois_JPA2016}.

    \item Define a set of orthonormal vectors at the initial sampled point $\vb{r}_0$. The first vector ($\vb{\hat{w}}_0$) points in the direction of the pseudovorticity $\boldsymbol{\omega}_0$ evaluated at the initial sampled point; this indicates the direction of the vortex line. The other two ($\vb{\hat{u}}_0$, $\vb{\hat{v}}_0$) are arbitrary vectors orthogonal to $\boldsymbol{\omega}_0$ ordered such that $\vb{\hat{u}}_0 \times \vb{\hat{v}}_0 = \vb{\hat{w}}_0$.

    \item Apply a Newton–Raphson method to find the closest zero of $\psi$ from the initial sampled point $\vb{r}_0$ in the plane perpendicular to the vortex line. The vortex core estimation $\vb{r}_n$ at the $n$-th iteration is obtained according to $\vb{r}_{n+1} = \vb{r}_n + (\Delta_u r_n) \vb{\hat{u}}_n + (\Delta_v r_n) \vb{\hat{v}}_n$, where
    \begin{align}
        \begin{bmatrix}
            \Delta_{u} r_n \\
            \Delta_{v} r_n \\
        \end{bmatrix} &= -
        \begin{bmatrix}
            \text{Re} \left( \partial_{u}^{(n)} \psi_n \right) & \text{Re} \left( \partial_{v}^{(n)} \psi_n \right) \\
            \text{Im} \left( \partial_{u}^{(n)} \psi_n \right) & \text{Im} \left( \partial_{v}^{(n)} \psi_n \right)
        \end{bmatrix}^{-1}
        \begin{bmatrix}
            \text{Re} \left( \psi_n \right) \\
            \text{Im} \left( \psi_n \right)
        \end{bmatrix}, \\
        &= \frac{1}{|\boldsymbol{\omega}_n|} \begin{bmatrix}
            \text{Im} \left( \psi^*_n \partial_v^{(n)} \psi_n \right) \\
            -\text{Im} \left( \psi^*_n \partial_u^{(n)} \psi_n \right)
        \end{bmatrix}.
    \end{align}
    Here $\psi_n$, and $\boldsymbol{\omega}_n$ are the corresponding fields evaluated at $\vb{r}_n$. We also define the directional derivatives $\partial_u^{(n)} =  \vb{\hat{u}}_n \cdot \nabla$, $\partial_v^{(n)} =  \vb{\hat{v}}_n \cdot \nabla$, and update the orthonormal vectors $\vb{\hat{u}}_n$, $\vb{\hat{v}}_n$, $\vb{\hat{w}}_n$ for each position $\vb{r}_n$. The iterations are continued until a vortex core is found for $|\psi| / \sqrt{\lambda \xi^3} < \Lambda$ with a small value $\Lambda$, for which $\Lambda = 10^{-13}$ is chosen. To find the values $\psi_n$ and $\nabla \psi_n$ at each iteration with subgrid resolution, we use Fourier-based interpolation directly in the MPS ansatz, which we detail in Appendix~\ref{sec:VortexTracking2}.

    \item To find the coordinates of the subsequent point in the vortex line, the following initial point is chosen as $\vb{r}_{n+1} = \vb{r}_n + \zeta \vb{\hat{w}}_n$ where $\zeta$ is the resolution of the vortex line. This evolution is performed so that the next sampled point follows the direction of the vortex line given by $\vb{\hat{w}}_n$ at $\vb{r}_n$. The addition is modulo $L$ to track lines across periodic boundaries.

    \item The previous steps are repeated until the last $\vb{r}_n$ and first $\vb{r}_0$ vortex core positions fullfill $|\vb{r}_n - \vb{r}_0| < \zeta$, indicating the closing of the vortex line.
\end{enumerate}

This method is iterated to obtain the complete set of vortex lines. To avoid double counting, Ref.~\cite{Villois_JPA2016} proposed using a binary array to track the positions of the vortex lines. In contrast, for the grid sizes used here, this is a significant memory toll for the method. To circumvent this overhead, the first vortex position found for a new vortex line is compared directly to the data of previously found vortex lines and is discarded accordingly.

\subsection{Fourier interpolation}\label{sec:VortexTracking2}

To efficiently recover the information of the wavefunction $\psi$ and its derivatives $\nabla\psi$ at an arbitrary position $\vb{r}_n$ (in the $n$-th iteration of the Newton-Raphson step from Appendix~\ref{sec:VortexTracking1}) with subgrid resolution, we use Fourier interpolation~\cite{Garcia-Ripoll_Quantum2021}. The procedure consists of transforming the wavefunction ($\psi$) to Fourier space ($\mathcal{F}\left[ \psi \right]$) and then applying the inverse transform ($\mathcal{F}^{-1}$) evaluated at $\vb{r}_n$ as follows:
\begin{align} 
    \psi(\vb{r_n}) &= \mathcal{F}^{-1} \left[ \mathcal{F}\left[ \psi \right]  \right]_{\vb{r}_n} \label{eqn:QFTInterpolation1}\\
    &=  \sum_{\vb{k}} \frac{1}{\sqrt{M}} e^{i \vb{k} \cdot \vb{r}_n} \mathcal{F}\left[ \psi \right]_{\vb{k}}, \label{eqn:QFTInterpolation2}
\end{align}
where $M$ is the total number of gridpoints.

In the quantics MPS formalism, to obtain $\mathcal{F}\left[ \psi \right]_{\vb{k}}$ we use the quantum Fourier transform (QFT), which corresponds to a low-rank MPO~\cite{Chen_PRXQ2023}. We use the 1D construction of the QFT from Ref.~\cite{Chen_PRXQ2023} for each spatial dimension and merge them with delta tensors as illustrated in Fig.~\ref{fig:MPS2DTo3D}. This procedure maps each component of $\vb{k}$ to the domain $[0,\pi 2^N/L) \cup [-\pi 2^N/L,0)$ going from positive to negative values from left to right. To facilitate writing functions and operators in $k$-space we apply an additional $X = \left(\begin{smallmatrix} 0 & 1 \\ 1 & 0 \end{smallmatrix}\right)$ gate at the largest length scale (in $k$-space) to change the domain as $\vb{k} \in [-\pi 2^N/L,\pi 2^N/L)^3$, restablishing the real-line order.

In addition, the exponential term $ M^{-1/2} e^{i \vb{k} \cdot \vb{r}_n}$ as a function of $\vb{k}$ can be represented by an MPS with $\chi_{\text{max}} = 1$~\cite{Khoromskij_CA2011}. To apply the quantics encoding of the exponential term in Fourier space, we use the following transformation for one spatial dimension:
\begin{align} \label{eqn:QFT-QTT}
    k^{(\alpha)}(q_1,q_2,\dots,q_N) = \frac{2\pi}{L} 2^N \sum_{l=1}^N \frac{q_l}{2^l} - \frac{\pi}{L} 2^N, \quad q_l = 0, 1.
\end{align}
In contrast to Eq.~\eqref{eqn:QTT}, the previous definition maps each component $k^{(\alpha)}$ ($k^{(x)}$, $k^{(y)}$ or $k^{(z)}$) of $\vb{k}$ to the domain $[-\pi 2^N/L, \pi 2^N/L)$ instead of $[0,L)$. Also, care must be taken in following the order of the qubits $q_l$ from the largest to the smallest length scales (in $k$-space) since the QFT inverts the qubit order~\cite{Chen_PRXQ2023}. The quantics MPS of the one-dimensional exponential term is obtained by substituting Eq.~\eqref{eqn:QFT-QTT} in $ (2^N)^{-1/2} e^{i k^{(\alpha)} \cdot r^{(\alpha)}_n}$ ($r^{(x)}_n$, $r^{(y)}_n$ or $r^{(z)}_n$). This corresponds to a product state of the form
\begin{align} \label{eqn:ExponentialQTT}
    \frac{1}{\sqrt{2^N}} \bra{e^{i k^{(\alpha)} \cdot r^{(\alpha)}_n}} = e^{- i \frac{\pi 2^N}{L} r^{(\alpha)}_n} \prod_{l = 1}^N \left( \frac{\bra{0}_l + e^{i \frac{ 2\pi 2^{N-l}}{L} r^{(\alpha)}_n}\bra{1}_l}{\sqrt{2}} \right),
\end{align}
where $\bra{0}_l$, $\bra{1}_l$ are the basis states of the $l$-th qubit from Eq.~\eqref{eqn:QFT-QTT}. This definition is applied to each spatial dimension. Then, the resulting MPSs are multiplied using the Hadamard product. Finally, the exponential MPS $M^{-1/2}\bra{e^{i \vb{k} \cdot \vb{r}_n}}$ is contracted with the quantics MPS in Fourier space $\ket{\mathcal{F}\left[ \psi \right]_{\vb{k}}}$ as shown in Eqs.~\eqref{eqn:QFTInterpolation1} and \eqref{eqn:QFTInterpolation2} to obtain the interpolated value $\psi(\vb{r}_n) = M^{-1/2} \braket{e^{i \vb{k} \cdot \vb{r}_n}}{\mathcal{F}\left[ \psi \right]_{\vb{k}}}$.

To obtain the interpolation of the derivatives we use the $k$-space representation given by $\mathcal{F}\left[ \nabla \psi \right]_{\vb{k}} = -i \vb{k} \mathcal{F}\left[ \psi \right]_{\vb{k}}$. Then, in addition to Eqs.~\eqref{eqn:QFTInterpolation1} and \eqref{eqn:QFTInterpolation2}, it is only necessary to multiply by the additional MPOs $-i\vb{k}$ (one for each component). To create each component of $-i\vb{k}$, we use the mapping from Eq.~\eqref{eqn:QFT-QTT} and interpret each binary variable as a number operator $q_l \to \hat{q_l} = \left(\begin{smallmatrix} 0 & 0 \\ 0 & 1 \end{smallmatrix}\right)$ for each qubit and spatial dimension. The resulting expression can be used to create an MPO efficiently~\cite{Ren_JCP2020, Chan_JCP2016, Hubig_PRB2017, Keller_JCP2015, Ehlers_PRB2017, Corbett_2025} and is guaranteed to have a small bond dimension of $\chi_{\text{max}} = 2$ since it is a linear function~\cite{Khoromskij_CA2011}. Since the algorithm requires the interpolation of derivatives using the QFT, we also use this tool during the vortex tracking method to obtain the quantics MPS of $\nabla \psi$ in position space. This calculation is performed as $\nabla \psi(\vb{r}) = \mathcal{F}^{-1} \left[ -i\vb{k} \mathcal{F}\left[ \psi \right]_{\vb{k}}  \right]_{\vb{r}}$, where $\mathcal{F}^{-1}$, $\mathcal{F}$ and $-i\vb{k}$ are low-rank MPOs while $\nabla \psi$ and $\psi$ are MPSs, for the components in each direction.

\subsection{Computational cost}\label{sec:VortexTracking3}

In analogy to the time evolution method presented in Sec.~\ref{sec:TimeEvolution}, the largest computational costs scale as $\mathcal{O}(\chi^4)$ and correspond to the Hadamard products used to calculate $|\boldsymbol{\omega}|^2$. Then, the calculation of the MPSs $\mathcal{F}\left[\psi\right]$ and $\mathcal{F}\left[\nabla \psi\right]$ each scales as $\mathcal{O}(\chi^3)$ and only has to be done once per snapshot. Each sampling costs $\mathcal{O}(\chi^2)$~\cite{Stoudenmire_NJP2010, Ferris_PRB2012} and each performed interpolation also costs $\mathcal{O}(\chi^2)$ since the bond dimension of the MPS to be contracted with is equal to one. The number of interpolations approximately scales with the vortex line length of the snapshot, as sampling with $|\boldsymbol{\omega}|^2$ ensures that the sampled points are close to the vortex lines. This last step can be efficiently parallelized, analogous to the original algorithm~\cite{Villois_JPA2016}, since the tracking of each separate vortex line can be done independently of the others, up to the point of double counting.

\section{Convergence of vortex reconnection simulation}\label{sec:ConvergenceReconnection}

\begin{figure}[t]
    \centering
    \includegraphics[width=\linewidth]{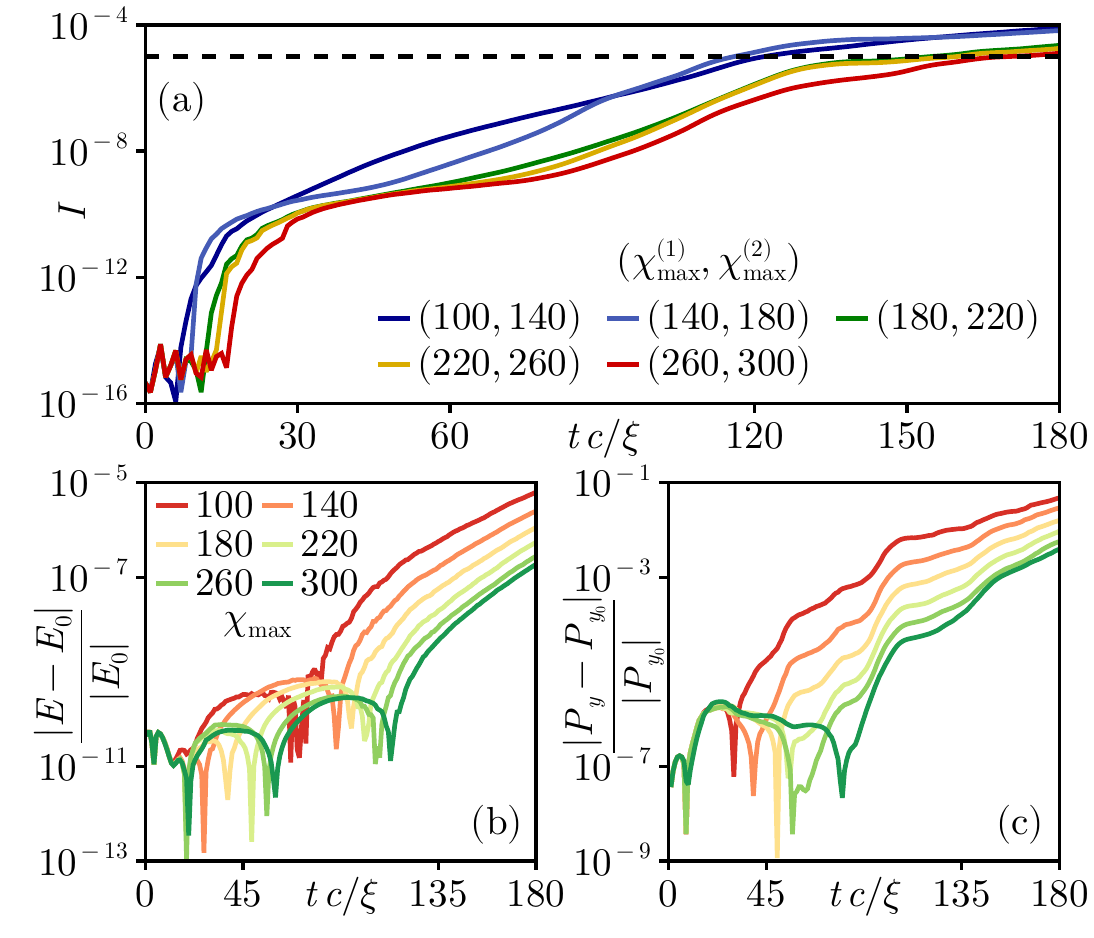}
    \caption{Convergence tests of the MPS results from Figs.~\ref{fig:VortexReconnectionInit}-\ref{fig:VortexReconnectionWaves}. (a) Infidelity for two consecutive values of $\chi_{\text{max}}$ as a function of time $t$. The bond dimensions are chosen so $\chi_{\text{max}}^{(2)} - \chi_{\text{max}}^{(1)} = 40$. The dashed line indicates $I = 10^{-5}$. (b,c) Relative change in energy $E$ (b) and linear momentum in the $y$ direction $P_y$ (c) with respect to the initial values $E_0$ and $P_{y_{0}}$, respectively. Both (b,c) have the same color coding.}
    \label{fig:VortexReconnectionInfidelity}
\end{figure}

To assess the validity of the vortex reconnection simulation of Sec.~\ref{sec:VortexLineReconnection} without resorting to a comparison with the DNS, we calculate the infidelity between two MPS simulations as a function of time. That is, we successively increase the bond dimension in steps of $\Delta \chi_{\text{max}} = 40$ up to a maximum value of $\chi_{\text{max}} = 300$. These results are presented in Fig.~\ref{fig:VortexReconnectionInfidelity}(a) where we identify three different regimes: early times $t \lesssim 20 \, \xi/c$, during the initial reconnection process signaled by low infidelities of $I \approx 10^{-15}$; intermediate times between $20 \, \xi/c \lesssim t \lesssim 90 \, \xi/c$, where the infidelities increase to $I \approx 10^{-9}$ due to the propagation of Kelvin waves and subsequent emission of vortex rings; and final times $t \gtrsim 90 \, \xi/c$, when the vortex dipole along the $x$-axis collapses into multiple vortex rings, resulting in an abrupt increase in the infidelity. For the simulation times considered in our calculation, the infidelity between $\chi_{\text{max}} = 260$ and $\chi_{\text{max}} = 300$ is capped at $I \leq 10^{-5}$ (dashed line). This indicates that the results do not vary significantly as $\chi_{\text{max}}$ increases, hence we use $\chi_{\text{max}} = 260$ for our calculations.

To further emphasize the validity of the results, we calculate the total energy $E$ and linear momentum $\vb{P} = (P_x, P_y, P_z)$ to verify that these quantities are conserved, as expected for the undamped GP equation. These measurements are defined as~\cite{Barenghi_2016}
\begin{align} 
    E &= \int \left[ \frac{\hbar^2}{2m} |\nabla \psi|^2 + V|\psi|^2 +\frac{g}{2}|\psi|^4 \right] d^3\vb{r}, \\
    \vb{P} &= \hbar \int \frac{\psi^* \nabla\psi - \psi \nabla \psi^*}{2i} d^3\vb{r}.
\end{align}
The advantage of the MPS framework is that both quantities can be calculated directly using inner products of the MPS representations of $\ket{\psi}$, $\ket{|\psi|^2}$, and $\ket{\nabla \psi}$ (per component). For example, the linear momentum corresponds to $\vb{P} = \hbar \, \text{Im}(\braket{\psi}{\nabla \psi})$ where the inner product is the contraction of both MPS with their joint free indices. The derivatives are calculated as discussed in Appendix~\ref{sec:Derivatives}. In Fig.~\ref{fig:VortexReconnectionInfidelity}(b,c), we show the relative change in energy $E$ and linear momentum in the $y$ direction $P_y$ with respect to their initial values $E_0$ and $P_{y_0}$ for different $\chi_{\text{max}}$ values. We observe that for $\chi_{\text{max}} = 260$ each relative error varies as $\approx 10^{-7}$ and $ \approx 10^{-3}$, respectively, showing that the quantities are well conserved. Since the initial movement of the vortex lines in Figs.~\ref{fig:VortexReconnectionInit}-\ref{fig:VortexReconnectionWaves} is only in the $y$ direction, we have $P_x = P_z = 0$. We have also verified that these values are conserved to within absolute errors of $ \approx 10^{-8}$.

\bibliography{QuantumTurbulence}

\end{document}